\documentclass[sigconf, nonacm]{acmart}

\usepackage{hyperref}
\usepackage{url}
\usepackage{bm}
\usepackage{amsmath,amsfonts}
\usepackage{graphicx}
\usepackage{textcomp}
\usepackage{xcolor}
\usepackage{multirow}
\usepackage{makecell}
\usepackage{subfigure}
\usepackage{color}
\usepackage{caption}
\usepackage{braket}

\usepackage[linesnumbered,ruled]{algorithm2e}

\newtheorem{myDef}{Definition}
\newtheorem{myTheorem}{Theorem}

\usepackage[utf8]{inputenc}
\usepackage{cleveref}
\crefname{section}{§}{§§}
\Crefname{section}{§}{§§}
\def\BibTeX{{\rm B\kern-.05em{\sc i\kern-.025em b}\kern-.08em
    T\kern-.1667em\lower.7ex\hbox{E}\kern-.125emX}}

\newcommand\vldbdoi{XX.XX/XXX.XX}
\newcommand\vldbpages{XXX-XXX}
\newcommand\vldbvolume{17}
\newcommand\vldbissue{1}
\newcommand\vldbyear{2024}
\newcommand\vldbauthors{\authors}
\newcommand\vldbtitle{\shorttitle} 
\newcommand\vldbavailabilityurl{URL_TO_YOUR_ARTIFACTS}
\newcommand\vldbpagestyle{plain} 

\begin{document}
\title{Routing-Guided Learned Product Quantization for Graph-Based Approximate Nearest Neighbor Search}

\author{Qiang Yue$^1$, Xiaoliang Xu$^1$, Yuxiang Wang$^1$, Yikun Tao$^1$, Xuliyuan Luo$^1$}
\affiliation{$^1$\institution{Hangzhou Dianzi University, China}}
\email{{yq,xxl,lsswyx,tyk,xuliyuanluo}@hdu.edu.cn}

\begin{abstract}
Given a vector dataset $\mathcal{X}$, a query vector $\vec{x}_q$, graph-based Approximate Nearest Neighbor Search (ANNS) aims to build a proximity graph (PG) as an index of $\mathcal{X}$ and approximately return vectors with minimum distances to $\vec{x}_q$ by searching over the PG index. It suffers from the large-scale $\mathcal{X}$ because a PG with full vectors is too large to fit into the memory, e.g., a billion-scale $\mathcal{X}$ in 128 dimensions would consume nearly 600 GB memory. To solve this, Product Quantization (PQ) integrated graph-based ANNS is proposed to reduce the memory usage, using smaller compact codes of quantized vectors in memory instead of the large original vectors. Existing PQ methods do not consider the important routing features of PG, resulting in low-quality quantized vectors that affect the ANNS's effectiveness. In this paper, we present an end-to-end Routing-guided learned Product Quantization (RPQ) for graph-based ANNS. It consists of (1) a \textit{differentiable quantizer} used to make the standard discrete PQ differentiable to suit for back-propagation of end-to-end learning, (2) a \textit{sampling-based feature extractor} used to extract neighborhood and routing features of a PG, and (3) a \textit{multi-feature joint training module} with two types of feature-aware losses to continuously optimize the differentiable quantizer. As a result, the inherent features of a PG would be embedded into the learned PQ, generating high-quality quantized vectors. Moreover, we integrate our RPQ with the state-of-the-art DiskANN and existing popular PGs to improve their performance. Comprehensive experiments on real-world large-scale datasets (from 1M to 1B) demonstrate RPQ's superiority, e.g., 1.7$\times$-4.2$\times$ improvement on QPS at the same recall@10 of 95\%.
\end{abstract}
%It has been widely recognized that graph-based ANNS is effective and efficient, however
% which easily can be adaptive to existing popular PGs that facilitate the search's performance.
%  replacing a large PG with original vectors by the one with smaller compact codes of quantized vectors
\maketitle

\pagestyle{\vldbpagestyle}
\begingroup\small\noindent\raggedright\textbf{PVLDB Reference Format:}\\
\vldbauthors. \vldbtitle. PVLDB, \vldbvolume(\vldbissue): \vldbpages, \vldbyear.\\
\href{https://doi.org/\vldbdoi}{doi:\vldbdoi}
\endgroup
\begingroup
\renewcommand\thefootnote{}\footnote{\noindent
This work is licensed under the Creative Commons BY-NC-ND 4.0 International License. Visit \url{https://creativecommons.org/licenses/by-nc-nd/4.0/} to view a copy of this license. For any use beyond those covered by this license, obtain permission by emailing \href{mailto:info@vldb.org}{info@vldb.org}. Copyright is held by the owner/author(s). Publication rights licensed to the VLDB Endowment. \\
\raggedright Proceedings of the VLDB Endowment, Vol. \vldbvolume, No. \vldbissue\ %
ISSN 2150-8097. \\
\href{https://doi.org/\vldbdoi}{doi:\vldbdoi} \\
}\addtocounter{footnote}{-1}\endgroup

\ifdefempty{\vldbavailabilityurl}{}{
\vspace{.3cm}
\begingroup\small\noindent\raggedright\textbf{PVLDB Artifact Availability:}\\
The source code, data, and/or other artifacts have been made available at \url{\vldbavailabilityurl}.
\endgroup
}

\section{Introduction}
\label{sec:intro}
\textit{Approximate Nearest Neighbor Search} (ANNS) ~\cite{arya1993approximate, indyk1998approximate, arya1998optimal} is a fundamental problem in many real-world applications, including recommendation systems ~\cite{wang2022fast, sarwar2001item, meng2020pmd, sarwar2001item}, information retrieval ~\cite{xu2022academic, flickner1995query, zhu2019accelerating}, data mining \cite{adeniyi2016automated, bijalwan2014knn, huang2017query}, and pattern recognition\cite{cover1967nearest, zhu2019accelerating, kosuge2019object, torralba2008small}. Given a vector dataset $\mathcal{X}$ and a query vector $\vec{x}_q$, ANNS efficiently and effectively returns relevant results to $\vec{x}_q$ from $\mathcal{X}$ with minimum distance. With the emergence of neural embedding \cite{amvrosiadis2019data, datta2008image, wang2015learning}, sparse discrete data (e.g., documents) can be transformed into dense continuous vectors. This renders ANNS an increasingly important problem, especially for information retrieval over large-scale vectorized datasets. Additionally, with the emergence and growing popularity of \textit{Large Language Models} (LLMs), ANNS is often applied to retrieve relevant information from external database for prompting LLMs \cite{dobson2023scaling}, consequently enhancing LLMs' reliability.

Among various types of ANNS methods, the graph-based ANNS is now a mainstream solution \cite{malkov2018efficient, chen2021spann, fu2017fast, liu2022optimizing}, with a vast amount of theoretical and empirical literature \cite{aumuller2020ann, Wang2021, shimomura2021survey} has proven their potential. Graph-based ANNS generally involves two phases: preprocessing phase and query processing phase. In the preprocessing phase, a \textit{Proximity Graph} (PG, we formally define it in \S \ref{sec:preliminaries}) is constructed as an index based on the vector dataset $\mathcal{X}$, each vertex in a PG corresponds a vector in $\mathcal{X}$, and an edge between two vertices denotes a neighbor relationship. During the query processing phase, a routing process is conducted over a PG. It starts from an entry vertex and iteratively exploring the PG along the vertices that can guide the search closer to the query $\vec{x}_q$ until it converges \cite{baranchuk2019learning}. With the excellent routing navigation ability of a PG, graph-based ANNS can quickly locate the neighbors that are close to optimal, which translates to a better trade-off between recall and latency \cite{aoyama2013graph, aumuller2020ann, hacid2010neighborhood}.

\begin{figure*}
\vspace{-0.1cm}
\centering
\includegraphics[width=0.92\linewidth]{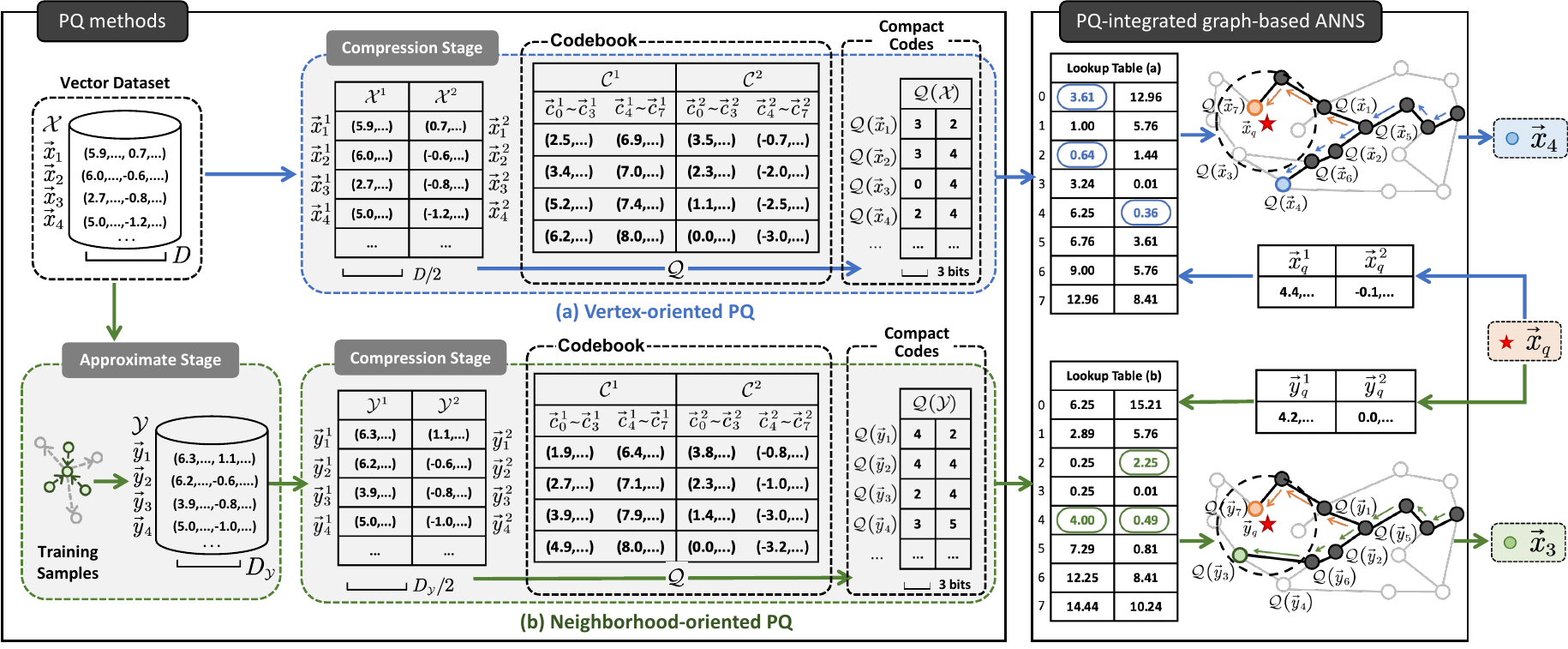}
\vspace{-0.2cm}
\caption{\textbf{Overview of PQ methods for graph-based ANNS. (a) Vertex-oriented PQ and (b) Neighborhood-oriented PQ.}}\label{fig:1}
\vspace{-0.5cm}
\end{figure*}

Although graph-based ANNS methods have been widely studied and achieve excellent performance, they face limitations when dealing with large-scale (million or even billion scale) vector datasets, which are commonly encountered in real-world scenarios \cite{arora2018hd, douze2018link}. This is because graph-based ANNS methods assume that a PG fits in the memory, which leads to a extreme large memory footprint. For example, using a graph-based ANNS method for a billion floating-point vectors in 128 dimensions would consume nearly 600 GB memory, far exceeding the RAM capacity of a workstation.

Various algorithms \cite{jayaram2019diskann, zhang2019grip, douze2018link, singh2021freshdiskann} have been proposed to combine with \textit{Product Quantization} (PQ) \cite{jegou2010product}, in order to augment graph-based ANNS for large-scale datasets. Specifically, taking DiskANN \cite{jayaram2019diskann} as an example, vectors are compressed by PQ, only the compact codes and codebook are cached in main memory for pre-computation, while the PG and original vectors are stored in external memory for post-verification. The quantization quality will significantly affect ANNS's effectiveness, e.g., reducing the memory overhead by 16$\times$ can result in a 25\% decrease in recall. This can be explained that quantization process does not carry the inherent features of a PG. \textit{This motivates us to explore a well-designed PQ that is tailored for PG, thus improving the graph-based ANNS's effectiveness as well as conserving memory usage}. 

\vspace{0.1cm}
\noindent\textbf{Existing PQ methods.} We generally categorize existing PQ methods into the following two categories.

\vspace{0.1cm}
\noindent\underline{Vertex-oriented PQ \cite{jayaram2019diskann, zhang2019grip, jegou2010product, ge2013optimized, norouzi2013cartesian}}. This line of work aims to minimize the distortions between the original vectors and the quantized vectors. Figure \ref{fig:1} (a) illustrates an example. A vector dataset $\mathcal{X}$ is given in the Euclidean space $\mathcal{E}^D$, where each vector $\vec{x}_i \in \mathcal{X}$ has $D$ dimensions ($D=16$ in Figure \ref{fig:1} (a)). To quantize each vector as a compact code, a quantizer $\mathcal{Q}$ is established based on $\mathcal{X}$ as follows: (1) $\mathcal{X}$ is divided into $M$ ($M=2$ in Figure \ref{fig:1}) non-overlapping chunks $\mathcal{X}=\{\mathcal{X}^1,\dots, \mathcal{X}^M\}$, where $\mathcal{X}^j$ is the $j$-th chunk of $\mathcal{X}$, and the sub-vector of vector $\vec{x}_i$ within $\mathcal{X}^j$ is denoted as $\vec{x}^{j}_i$; (2) A clustering algorithm (e.g. $k$-means) is applied to each chunk to generate $K$ ($K=8$ in Figure \ref{fig:1}) clusters. The centroid of each cluster in $\mathcal{X}^j$ is denoted as a codeword $\vec{c}^{\, j}_k$, where $k=\{0,\dots,K-1\}$ is the identifier of the codeword. The $K$ codewords of each chunk $\mathcal{X}^j$ form the sub-codebook $\mathcal{C}^j=\{\vec{c}_{0}^{\,j},\dots, \vec{c}_{K-1}^{\,j}\}$. The codebook of $\mathcal{X}$ is defined as the Cartesian product of sub-codebooks, denoted as $\mathcal{C} = \mathcal{C}^1 \times \dots \times \mathcal{C}^M$; (3) The Lloyd \cite{1056489, jegou2010product} quantizer $\mathcal{Q}$ is often established to encode vector $\vec{x}_i$ as a compact code $\mathcal{Q}(\vec{x}_i)$. Specifically, $\mathcal{Q}$ finds the closest codeword w.r.t. each sub-vector $\vec{x}^j_i$ in $\mathcal{X}^j$ and quantizes $\vec{x}_i$ using the identifiers of $M$ closest codewords as its compact code $\mathcal{Q}(\vec{x}_i)$. For example, consider the input vector $\vec{x}_1$, which is divided into two sub-vectors $\{\vec{x}^1_1,\vec{x}^2_1\}$. Suppose that $\vec{x}^1_1$ is close to the fourth codeword $\vec{c}_3^{\,1}$ in $\mathcal{X}^1$ and $\vec{x}^2_1$ is close to $\vec{c}_2^{\,2}$ in $\mathcal{X}^2$, then we have the compact code $\mathcal{Q}(\vec{x}_1)=\{3,2\}$. Since a codeword can be identified by an integer using $\log_2K$ bits, in this example, each compact code only requires $M\log_2K=6$ bits for storage, which is much less than that of original vector, i.e., $D * \text{sizeof(float)} = 64$ bytes. It is clear to observe that using the compact code $\mathcal{Q}(\mathcal{X})$ would significantly reduce the memory overhead of maintaining full vectors.

In querying phase, given a query vector $\vec{x}_q$, it is initially vertically divided into $M$ sub-vectors $\vec{x}_q=\{\vec{x}_q^1,\dots, \vec{x}_q^M\}$. Next, a lookup table is pre-computed to cache the distances between each $\vec{x}_q^j$ and every codeword $\vec{c}^{\,j}_{k}$ in the sub-codebook $\mathcal{C}^j$. In this way, the distance between $\vec{x}_q$ and any $\vec{x}_i\in \mathcal{X}$ can be estimated by the one between $\vec{x}_q$ and $\mathcal{Q}({\vec{x}_i})$. Given a lookup table (a), assume that we want to estimate the distance between $\vec{x}_q$ and $\vec{x}_3$, we only require to retrieve the entries from lookup table (a) using the compact code $\mathcal{Q}(\vec{x}_3)=\{0,4\}$, i.e., the distances of $\vec{x}_q^{\,1}$ and $\vec{x}_q^{\,2}$ to $\vec{c}^{\,1}_0$ and $\vec{c}^{\,2}_4$ that are 3.61 and 0.36, and then estimate the distance as 3.61+0.36.

Since vertex-oriented PQ does not consider the neighborhood relationships between nodes in a PG, it suffers from the inaccurate ANNS results. In Figure \ref{fig:1}, following the blue routing path we reach to $\vec{x}_6$ and consider its neighbors $\vec{x}_3$ and $\vec{x}_4$ as candidates for next-hop selection. Note that, in the neighborhood of $\vec{x}_q$ (dotted circle), $\vec{x}_3$ is closer to the query $\vec{x}_q$ than $\vec{x}_4$. However, using the lookup table (a) would mistakenly select $\vec{x}_4$ as next-hop as it's quantized vector is closer to $\vec{x}_q$ (i.e., 0.64+0.36$<$3.61+0.36). Thus, the ranking of candidates for next-hop selection becomes distorted, leading to the failure of PG routing to converge. \textit{This example demonstrates that maintaining neighborhood relationships of a PG is essential for establishing a good quantizer $Q$ for graph-based ANNS}.

\vspace{0.1cm}
\noindent\underline{Neighborhood-oriented PQ \cite{zhang2022connecting, karaman2019unsupervised, sablayrolles2018spreading, prokhorenkova2020graph}} This line of work incorporates neighborhood relationships into its approach following two steps as shown in Figure \ref{fig:1} (b). First, it learns the approximate vectors $\mathcal{Y}$ for original vectors in $\mathcal{X}$, ensuring that vertices in the same neighborhood are more similar in $\mathcal{Y}$ than others. Second, it compresses $\mathcal{Y}$ into compact codes $\mathcal{Q}(\mathcal{Y})$ by the same produce as vertex-oriented PQ. It still faces limitations when candidates do not belong to the neighborhood of $\vec{x}_q$. Considering $\vec{y}_5$ in the green routing path, its two neighbors $\vec{y}_1$, $\vec{y}_2$ are out of $\vec{x}_q$'s neighborhood. Using the lookup table (b) would mistakenly select $\vec{y}_2$ as next-hop because 4.00+0.49$<$4.00+2.25, leading to a final result as $\vec{y}_3$ not the optimal $\vec{y}_7$. \textit{This example demonstrates that, besides preserving the neighborhood relationships in $\mathcal{Q}$, maintaining the important routing features (i.e., correct next-hop selection during the routing procedure)} is also essential for graph-based ANNS. If we do so, the PQ with routing features would correctly select $\vec{y}_1$ to guide a search located in $\vec{x}_q$'s neighborhood and find the optimal results.
%Although it can distinguish that $\vec{x}_3$ is closer to $\vec{x}_q$ than $\vec{x}_4$,

\vspace{0.1cm}
\noindent \textbf{Challenges and our solutions.} Different from the aforementioned methods, we expect to establish a quantizer $\mathcal{Q}$ considering the features from both the neighborhood relationship in a PG and the routing process performed over a PG. Intuitively, one can come up with a straightforward method following the same two-step process as the typical neighborhood-oriented PQ \cite{prokhorenkova2020graph, sablayrolles2018spreading, zhang2022connecting}: It first embeds two types of features into an intermediate approximate vectors $\mathcal{Y}$, then it compresses $\mathcal{Y}$ into compact codes $\mathcal{Q}(\mathcal{Y})$ using existing PQ methods. Since the PQ procedure is not contained in the learning model of transforming $\mathcal{X}$ to $\mathcal{Y}$, the embedded features in the first step would be largely lost in the second step. More precisely, a two-step process would destroy the well-trained embeddings to some extent. This motivates us to study a one-segment end-to-end \textbf{R}outing-guided \textbf{P}roduct \textbf{Q}uantization (RPQ) for graph-based ANNS, which is non-trivial because of the following challenges.
%Consequently, it is crucial to develop a one-segment quantizer instead of a two-step one. 

%In \textbf{\S \ref{sec:model}}, we present a differentiable quantizer with two major steps: (1) Adaptive vector decomposition based on space rotation using a square orthonormal matrix. In this way, we can automatically assign dimensions to each sub-vectors so that each sub-vector would be equalized valuable for the following quantization step; (2) Differentiable quantization for above balanced sub-vectors, based on a differentiable approximation with codeword assignment probability and Gumbel-Softmax. In this way, we can generate approximate compact code in a continuous space instead of a discrete space, thus making the back-propagation possible. As a result, a PG's inherent features would be retained lossless in the learned PQ.

\vspace{0.1cm}
\noindent\underline{Challenge I:} \textit{How to make a discrete quantization process differentiable, enabling back-propagation for end-to-end learning of RPQ?} Since PQ requires vertically dividing each vector $\vec{x}_i\in \mathcal{X}$ into $M$ sub-vectors, the dimensions with valuable intrinsic features would unbalancedly locate among $M$ sub-vectors, resulting in meaningless quantized sub-vectors from those sub-vectors with less intrinsic features \cite{li2019approximate, amsaleg2015estimating, he2012difficulty}. Unfortunately, we cannot utilize a back-propagation to automatically optimize the dimensions' distribution in sub-vectors because the vertical division is non-differentiable. Besides, PQ assigns an identifier of the closest codeword to each sub-vector as its compact code, this is a non-differentiable \textit{argmin} operation. We cannot utilize back-propagation to optimize the compact code generation. In a nutshell, the key of end-to-end RPQ is to convert a non-differentiable quantization into a differentiable one, so that we can optimize a learned quantizer $\mathcal{Q}$ via back-propagation.

To handle this, in \textbf{\S \ref{sec:model}}, we present a differentiable quantizer $\mathcal{Q}$ with two major steps: (1) \textit{Adaptive vector decomposition} based on space rotation using a square orthonormal matrix. Since the space rotation is a differentiable vector operation, we can use back-propagation to update the square orthonormal matrix so as to refine the dimensions' districution among all sub-vectors, i.e., changing the vertical division to an automatic vector decomposition. (2) \textit{Differentiable quantization} based on an approximate codeword assignment probability calculated by the differentiable Gumbel-Softmax. In this step, we obtain an approximate compact code in a continuous space instead of a discrete space, making the back-propagation possible. As a result, the entire PQ procedure can be continuously updated via learning with back-propagation and a PG's inherent features would be retained lossless in the learned PQ.

\vspace{0.1cm}
\noindent\underline{Challenge II:} \textit{How to effectively extract a PG's neighborhood and routing features that are beneficial to the learning of RPQ?} A good quantizer should be able to make the quantized vectors of two vertices in a PG reflect their original neighborhood relationship. If two vertices are close in a PG, then their quantized vectors should be close too. The difficulty is how to define positive/negative samples w.r.t. neighborhood relationship and control their proportion to deal with the hard sample issue (i.e., distinguish two vertices with close distance but not neighborly) \cite{Cohan2020}. For routing features, a simple method is to use the optimal routing paths of training queries to predict routing paths for testing queries. However, it is problematic because in real-applications, queries are dynamically changing so that the optimal routing path of one already seen query is different from other unseen queries. Therefore, comparing with learning entire routing paths, it is more valuable to learn the decision making process (i.e., next-hop selection) of graph-based ANNS.

%we argue that learning the decision making process is more valuable for graph-based ANNS.
%A possible approach is to follow the idea of comparative learning to collect positive and negative samples w.r.t. neighborhood relationship, but the difficulty is \textit{how to define positive/negative samples and control their proportion to deal with the hard sample issue (i.e., distinguish two vertices with close distance but not neighborly) \cite{Cohan2020}}.
%we employ the idea of contrastive learning to collect positive/negative samples w.r.t. neighborhood relationship. To achieve this, we introduce two parameters $k_{\rm pos}$ and $k_{\rm neg}$ to define positive and negative samples of a vertex $v$ as its top-$k_{\rm pos}$ nearest vertices from the $n$-hop neighborhood of $v$ (in a PG), denoted by $N_n(v)$, and the top-$k_{\rm neg}$ vertices from the remained $N_n(v)-k_{\rm pos}$ vertices. In this way, we can flexibly control the proportion of positive and negative samples and the scope of hard samples by adjusting two parameters.

In \textbf{\S \ref{sec:loss}}, {we employ the idea of contrastive learning to collect positive and negative samples of a vertex $v$ in a PG as its top-$k_{\rm pos}$ nearest vertices from the $n$-hop neighborhood of $v$, denoted by $N_n(v)$, and the top-$k_{\rm neg}$ vertices from the remained $N_n(v)-k_{\rm pos}$ vertices, respectively. We can flexibly control the proportion of positive and negative samples and the scope of hard samples by adjusting two parameters $k_{\rm pos}$ and $k_{\rm neg}$. For routing features, we randomly select a set of queries and perform graph-based ANNS using the learned quantizer $\mathcal{Q}$. Then, we record all the ranked candidates used for next-hop selection in the entire routing path as routing features. We expect to learn how to select the correct next-hop from the ranked candidates, thus optimizing the learned $\mathcal{Q}$.

%by continuously optimize the learned $\mathcal{Q}$

\vspace{0.1cm}
\noindent\underline{Challenge III:} \textit{How to design an appropriate loss function to optimize the learned quantizer $\mathcal{Q}$ via back-propagation?} Since we have two different features considered to optimize the learned quantizer $\mathcal{Q}$, it is reasonable to design a dedicated loss function for each of them to optimize $\mathcal{Q}$ separately. However, this does not necessarily ensure the consistency of the gradient direction of two losses (i.e., the optimization direction may deviate). This implies us to integrate both losses to optimize $\mathcal{Q}$ along a unified gradient direction.

In \textbf{\S \ref{sec:training}}, we first present a neighborhood feature-aware loss to preserve the neighborhood relationships in the quantized vector space as that exists in the original vector space. Next, we propose a routing feature-aware loss to measure how good a decision made based on learned $\mathcal{Q}$, and minimize it to optimize $\mathcal{Q}$. Finally, we apply a multi-feature joint loss of the above two losses to optimize $\mathcal{Q}$ using mini-batch gradient, so as to unify the optimization direction.

\vspace{0.1cm}
\noindent \textbf{Contributions.} Our contributions are summarized as follows:
\begin{itemize}
    \item We introduce an end-to-end RPQ framework in \S \ref{sec:roq}, which is designed to be adaptive to existing popular PGs and facilitate PQ-integrated graph-based ANNS.
    \item We propose a differentiable quantizer in \S \ref{sec:model} to enable the back-propagation for end-to-end learning of RPQ.
    \item We present a sampling-based method to extract a PG's inherent features (\S \ref{sec:loss}), including the important neighborhood relationship features and routing features. 
    \item In \S \ref{sec:training}, we present a multi-feature joint training with two feature-aware losses to optimize the differentiable quantizer.
    \item Extensive experiments on real-world datasets (\S \ref{sec:experiments}) demonstrate the effectiveness and efficiency of RPQ.
\end{itemize}

%using the above two types of features.
%We propose a differentiable quantizer in \S \ref{sec:model} with two steps of adaptive vector decomposition and differentiable quantization, which enables the back-propagation of end-to-end learning of RPQ.
%In \S \ref{sec:loss}, we present a sampling-based method to extract a PG's inherent features, including the important routing features as well as the neighborhood features. A multi-joint training with two feature-aware losses are conducted to optimize the differentiable quantizer.

\vspace{-0.25cm}
\section{PRELIMINARIES AND PROBLEMS} \label{sec:preliminaries}

We show preliminaries in \S \ref{sec:pre} and define the problem in \S \ref{sec:problem}. Frequently used notations are provided in Table \ref{tab:freq}.

\setlength{\textfloatsep}{0.2cm}
\begin{table}
  \setlength{\abovecaptionskip}{0.1cm}
  %\captionsetup{labelfont=bf}
  \caption{Frequently used notations}
  \label{tab:freq}
  \renewcommand\arraystretch{0.92}
  \begin{tabular}{c|l}
    \hline
    \small{\textbf{Notations}}       &  \small{\textbf{Descriptions}}      \\
    \hline
    \hline
    \thead[c]{$\mathcal{X}$}            & \thead[l]{A limited dataset, where every element is a vector $\vec{x}_i$.} \\
    \hline
    \thead[c]{$\mathcal{X}^j$}          & \thead[l]{The $j$-th chunk of the dataset $\mathcal{X}$.} \\
    \hline
    \thead[c]{$\vec{x}_q$}              & \thead[l]{The query vector.} \\
    \hline
    \thead[c]{$\mathcal{C}$}            & \thead[l]{The codebook which is composed by sub-codebooks $\mathcal{C}^j$.} \\
    \hline
    \thead[c]{$\vec{c}_{k}^{\,j}$}      & \thead[l]{The $k$-th codeword of sub-codebook $\mathcal{C}^j$.} \\
    \hline
    \thead[c]{$\mathcal{Q}$}            & \thead[l]{A quantizer maps a vector $\vec{x}_i$ to a compact code $\mathcal{Q}(\vec{x}_i)$.} \\
    \hline
    \thead[c]{$\vec{x}_i'$}             & \thead[l]{A quantized vector of $\vec{x}_i$, which consists of codewords \\corresponding to $\mathcal{Q}(\vec{x}_i)$.} \\
    \hline
    \thead[c]{$\delta(\cdot, \cdot)$}								& {\thead[l]{The squared Euclidean distance \cite{jegou2010product} between \\two vectors.}} \\
    \hline
    \thead[c]{$G(V,E)$}                 & \thead[l]{A PG $G$ consists of vertices $V$ and edges $E$.} \\
    \hline
    %$v_i$                    & \textcolor{blue}{\thead[l]{The i-th vertex of vertices $V$. In PQ-integrated graph-based \\ANNS, each vertex represents a vector $\vec{x}_{v_i}$, a compact \\code $\mathcal{Q}(\vec{x}_{v_i})$ and a quantized vector $\vec{x}_{v_i}'$.} } \\
   % \hline
    \thead[c]{$N(v)$}                   & \thead[l]{The neighbors of vertex $v$ in $G$.} \\   
    \hline
\end{tabular}
\end{table}

%\textcolor{blue}{Unless otherwise specified, when referring to the PG's structure without the corresponding data, $v_i$ is used to describe. However, when discussing specific data types (e.g., $\vec{x}_i$, $\mathcal{Q}(\vec{x}_i)$ and $\vec{x}_i$), use the corresponding notation.}

\vspace{-0.15cm}
\subsection{Preliminaries}
\label{sec:pre}
\begin{myDef}
\label{def:anns}
\textbf{Approximate Nearest Neighbor Search (ANNS)} \cite{arya1993approximate, indyk1998approximate}. Given a vector dataset $\mathcal{X}$ of $n$ vectors, a query vector $\vec{x}_q$, and a parameter $\epsilon>0$. ANNS aims at effectively obtaining a vector $\vec{x}_p \in \mathcal{X}$ that is close to $\vec{x}_q$, satisfying the condition:
$\delta(\vec{x}_p,\vec{x}_q) \leq (1+\epsilon) \delta(\vec{x}_r, \vec{x}_q)$, where $\vec{x}_r$ is the nearest vector to $\vec{x}_q$ in $\mathcal{X}$.
\end{myDef}

\vspace{-0.1cm}
This problem can easily generalize to the case where we aim to return $k>1$ closest vectors to $\vec{x}_q$. In practice, we use \emph{recall@$k$} instead of the predefined $\epsilon$ to evaluate the accuracy. Given a $\vec{x}_q$, we use $\mathcal{R}$ to record $k$ vectors obtained by ANNS, and we use $\widetilde{\mathcal{R}}$ to record $k$ nearest vectors of $\vec{x}_q$. Then we define \emph{recall@$k$} as follows.
\begin{equation}
\label{eq:recall}
    Recall@k=\frac{|\mathcal{R} \cap \widetilde{\mathcal{R}}|}{|\widetilde{\mathcal{R}}|}=\frac{|\mathcal{R} \cap \widetilde{\mathcal{R}}|}{k}
\end{equation}

%The key of graph-based ANNS is the Proximity Graph (PG) that is defined below.
%Graph-based ANNS has gained prominence due to advantageous trade-off between accuracy and efficiency. It usually relay on the Proximity Graph (PG) defined as follows.

\begin{myDef}
\label{def:pg}
\textbf{Proximity Graph (PG) \cite{Wang2022NHQ,Wang2024}.} Given a vector dataset $\mathcal{X}$ and a non-negative distance threshold $\Theta$, we define the PG of $\mathcal{X}$ w.r.t. $\Theta$ as a graph $G = (V, E)$ with the vertex set $V$ and edge set $E$. (1) For each vector $\vec{x}_i\in \mathcal{X}$, it has a corresponding vertex $v_i\in V$. (2) For any two vertices $v_i$, $v_j\in V$, if $v_iv_j\in E$, we have $\delta(\vec{x}_{i}, \vec{x}_{j}) \leq \Theta$.
\end{myDef}

\vspace{0.1cm}

Similar to \cite{jegou2010product, faissweb, ge2013optimized}, we adopt the squared Euclidean distance between two vectors as $\delta(\cdot,\cdot)$, because it avoids square root operation so that it is more efficient than Euclidean distance. Graph-based ANNS assumes that both the PG and $\mathcal{X}$ would fit in main memory, resulting in a large memory footprint. One crucial optimization is using Product Quantization (PQ) to cut down the size of vectors. %We define PQ as below.

\vspace{0.1cm}

\begin{myDef}
\label{def:pq}
\textbf{Product Quantization (PQ).} Given a vector dataset $\mathcal{X}$ in $D$-dimensional space $\mathcal{E}^D$, and non-negative integers $M$ and $K$. PQ utilizes a quantizer $\mathcal{Q}$ to map a vector $\vec{x}_i \in \mathcal{X}$ to a compact code $\mathcal{Q}(\vec{x}_i)=\{\mathcal{Q}^1(\vec{x}_{i}^1),\dots,\mathcal{Q}^M(\vec{x}_{i}^M)\}$. (1) Each $\vec{x}_i^{\, j} \in \mathcal{E}^{D/M}$ is a sub-vector of the original $\vec{x}_i$. (2) Each sub-quantizer $\mathcal{Q}^j$ maps a sub-vector $\vec{x}_i^{\, j}$ to an identifier of a codeword $\vec{c}_k^{\,j}$ in sub-codebook $\mathcal{C}^j$. The mapping between sub-vector $\vec{x}_i^j$ and the codeword $\vec{c}_k^{\,j}$ is denoted as $\vec{c}_k^{\,j}=\mathcal{C}^j(\mathcal{Q}^j(\vec{x}_i^j))$. (3) Each sub-codebook $\mathcal{C}^j=\{\vec{c}_1^{\,j},\dots,\vec{c}_K^{\,j}\}$ is a set of $K$ codewords, and the codebook $\mathcal{C}$ is defined as the Cartesian product of sub-codebooks, i.e., $\mathcal{C} = \mathcal{C}^1 \times \dots \times \mathcal{C}^M$. Therefore, original vector $\vec{x}_i$ can be approximated by quantized vector $\vec{x}_i'=\mathcal{C}(\mathcal{Q}(\vec{x}_i))$.
\end{myDef}

\vspace{0.1cm}
 Given a vector dataset $\mathcal{X}$ and a vertex set $V$ of a PG, there is a bijection: $\mathcal{X}\rightarrow V$, that is, every vertex $v_i\in V$ is the image of exactly one vector $\vec{x}_i\in \mathcal{X}$. It's worth mentioning that when we need to emphasize the bijection between $\mathcal{X}$ and $V$, we will use $\vec{x}_{v_i}$ to clearly represent the corresponding vector of vertex $v_i$.

%(3) When we mention the vector distance $\delta(\cdot,\cdot)$ in the rest of this paper, it refers to the squared Euclidean distance calculated using the original vectors or quantized vectors (discussed later in \S \ref{sec:intuition}).}
%\noindent\textcolor{blue}{\textbf{Remarks.} (1) Given a vector dataset $\mathcal{X}$ and a vertex set $V$ of a PG, there is a bijection: $\mathcal{X}\rightarrow V$, that is, every vertex $v_i\in V$ is the image of exactly one vector $\vec{x}_i\in \mathcal{X}$. (2) Each vector $\vec{x}_i$ has a quantized vector $\vec{x}_i'$ using a quantizer $\mathcal{Q}$.}
%(3) When we mention the vector distance $\delta(\cdot,\cdot)$ in the rest of this paper, it refers to the squared Euclidean distance calculated using the original vectors or quantized vectors (discussed later in \S \ref{sec:intuition}).}

\subsection{Problem Definition}
\label{sec:problem}
Given the aforementioned definitions, we study the problem of designing a tailored PQ for effective graph-based ANNS.%, which defined below.

%we address critical shortcomings of current methods are incompatible with PG. We are focus on the following problem: improving both the accuracy and the retrieval performance by making the quantizer suit for global routing process of PG. We define this problem as follow.

\vspace{1ex}

%\textbf{Problem 1:} \emph{Given a vector dataset $\mathcal{X}$, a training query set $\mathcal{X}_{\rm train}$, a pre-built PG $G(V,E)$, and a result set $\mathcal{R_{\rm train}} \subset \mathcal{X}$ produced by performing ANNS over $G$ for all queries from $\mathcal{X}_{\rm train}$. We aim to obtain a quantizer $\mathcal{Q}^*$ that satisfies:}

\noindent\textbf{Problem:} \emph{Given a vector dataset $\mathcal{X}$, a query set $\mathcal{X}_{\rm query}=\{\vec{x}_{q_1},\dots,\vec{x}_{q_n}\}$ for $\mathcal{X}$, a pre-built PG $G(V,E)$, and a result set for all queries $\mathcal{R_{\rm query}}=\{\mathcal{R}_{1},\cdots,\mathcal{R}_n\}$ where $\mathcal{R}_i\subset \mathcal{X}$ contains $k$ vectors that are returned by performing ANNS over $G$ for each query vector $\vec{x}_{q_i}\in \mathcal{X}_{\rm query}$. We aim to obtain a quantizer $\mathcal{Q}$ that satisfies:}
%\vspace{-0.25cm}
\begin{equation}
%\small
%\fontsize{9.5pt}{2.6mm}\selectfont
\begin{split}
    \mathcal{Q} = arg \min_{\mathcal{Q}} \sum_{\vec{x}_{q_i} \in \mathcal{X}_{\rm query}} \sum_{\vec{x}_r \in \mathcal{R}_{i}} \delta((\mathcal{C}(\mathcal{Q}(\vec{x}_r)), \vec{x}_{q_i}) \quad .
\end{split}
\end{equation}

\emph{According to Definition \ref{def:anns}, $\mathcal{R}_{\rm query}$ satisfies:}
\begin{equation}
    \mathcal{R}_{\rm query} = \{arg \min_{\mathcal{R}_i \subset \mathcal{X}} \sum \limits_{\vec{x}_r \in \mathcal{R}_i} \delta(\vec{x}_r,\vec{x}_{q_i})|\forall \vec{x}_{q_i}\in \mathcal{X}_{\rm query}\}\quad .
\end{equation}

Intuitively, the smaller distance between the quantized vectors of all elements in $\mathcal{R}_i$ and $\vec{x}_{q_i}$, the better $\mathcal{Q}$ fits in ANNS over a PG.
%\noindent \emph{where the smaller distance between the quantized vectors of $\mathcal{R}$ and $\vec{x}_q$, the better $\mathcal{Q}$ fits in the retrieval process over a PG.}

\vspace{0.1cm}
\section{RPQ FRAMEWORK} \label{sec:roq}
We propose an end-to-end routing-guided product quantization (RPQ) for graph-based ANNS. RPQ combines the local neighborhood features and global routing features of a PG into a learned quantizer $\mathcal{Q}$. We start with the motivation behind our solution (\S \ref{sec:intuition}), then briefly introduce each component of RPQ in \S \ref{sec:framework}.

\begin{figure*}[t]
\vspace{-0.2cm}
\centering
\includegraphics[width=0.9\linewidth]{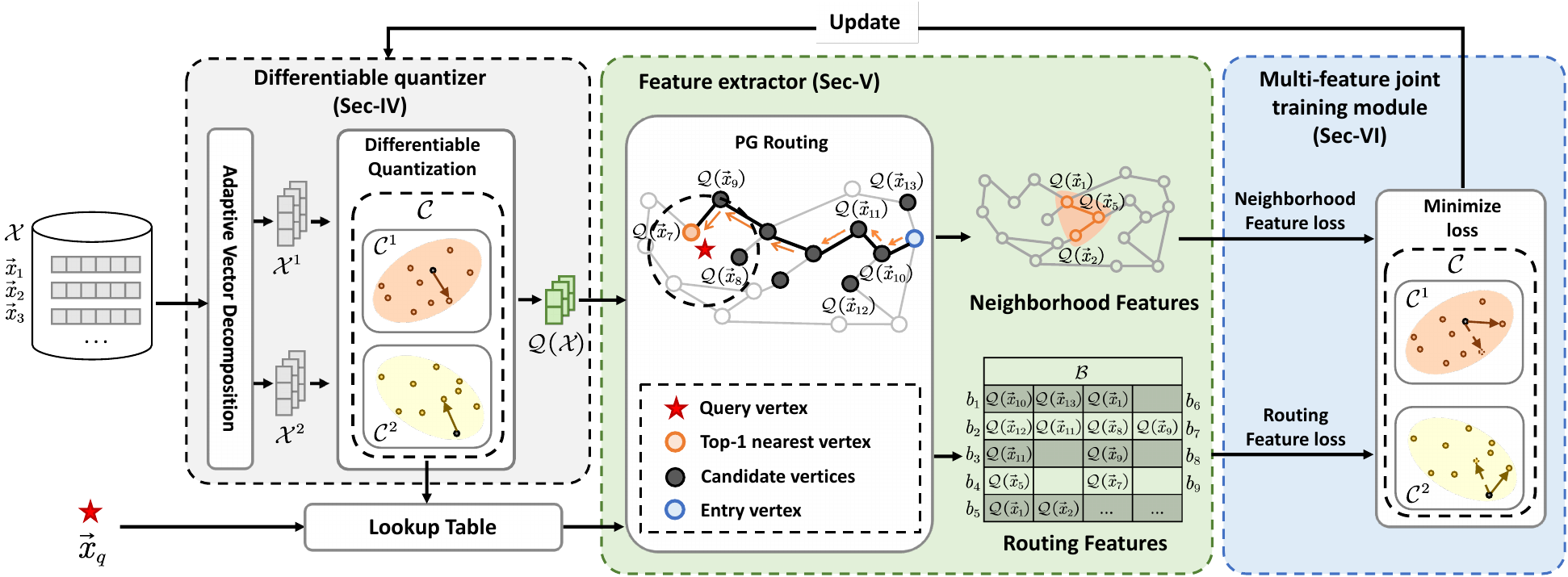}
\vspace{-0.3cm}
\caption{\textbf{A pipeline of our RPQ framework}}\label{fig:2}
\vspace{-0.4cm}
\end{figure*}

\subsection{Motivation} \label{sec:intuition}
Given a query vector $\vec{x}_q$ and a PG $G$, routing on $G$ is often performed as a \textit{beam search} \cite{prokhorenkova2020graph} that starts from an entry vertex and ends up at the $k$ vertices whose vectors are the closest to $\vec{x}_q$. The key of beam search is the next-hop selection at each visiting vertex. It maintains a global candidate set with $h$ candidate vertices ($h$ is the beam size that controls the search's width). During the beam search, the global candidate set is updated using each visiting vertex's neighbors based on their distances to $\vec{x}_q$ and the closest one would be selected as the next-hop. In PQ-integrated graph-based ANNS, such a distance can be computed in two ways: \textit{Symmetric Distance Computation} (SDC) \cite{jegou2010product} quantizes both $\vec{x}_q$ and $\vec{x}_v$ as $\vec{x}_q'$ and $\vec{x}_v'$, then computes distance as $\delta(\vec{x}_v',\vec{x}_q')$; \textit{Asymmetric Distance Computation} (ADC) \cite{jegou2010product} only quantizes $\vec{x}_v$ and computes distance as $\delta(\vec{x}_v',\vec{x}_q)$. ADC is widely used in practice as it yields a lower distance error and results in a better recall \cite{jegou2010product}. In this paper, we adopt ADC for distance computation. Given this premise, we next provide Theorem \ref{th:motivation} to clarify the importance of considering both neighborhood and routing features in RPQ for learning a good quantizer $\mathcal{Q}$.

\vspace{0.1cm}
\begin{myTheorem}
\label{th:motivation} Given a PG $G(V,E)$ and a query $\vec{x}_q$. Suppose the graph-based ANNS is now visiting a vertex $v\in V$, then the next-hop selection for ANNS at $v$ depends on the distances among $v$'s neighbors and the distances between all neighbors and $\vec{x}_q$.
%and it has $h$ candidates for search expansion
%\textcolor{red}{In the context of graph-based ANNS, given a PG $G(V,E)$ built for a vector dataset $\mathcal{X}$ and a query $\vec{x}_q$. Suppose that vertex $v \in V$ is currently being visited, let $v_x$ represent one of the neighbors of $v$, let $v_y$ denote any other neighbor. The selection of next-hop for ANNS at $v$ depends on both the distance between neighbors $\delta(\vec{x}_{v_x},\vec{x}_{v_y})$ and the distances between neighbors and query, i.e., $\delta(\vec{x}_{v_x}, \vec{x}_q)$ and $\delta(\vec{x}_{v_y}, \vec{x}_q)$.}
\end{myTheorem}

\vspace{-0.15cm}
\begin{proof}
The key of next-hop selection at a vertex $v$ is updating the global candidates of beam search using $v$'s neighbors. Given a neighbor $v_x\in N(v)$, its ranking order is determined by the distance comparison of the quantized vectors of $v_x$ and any other neighbor $v_y\in N(v)$ to $\vec{x}_q$, i.e., $\sum_{\vec{x}_{v_y} \in N(v)} \delta(\vec{x}_{v_x}',\vec{x}_{q}) -\delta(\vec{x}_{v_y}', \vec{x}_{q})$. 

We have the distance $\delta(\vec{x}_{v_x}',\vec{x}_{q})$ ($\delta(\vec{x}_{v_y}',\vec{x}_{q})$ is similar) as follows.
\begin{equation}
\hspace{-0.6cm}
\small
\begin{aligned}
    \delta(\vec{x}_{v_x}',\vec{x}_{q}) &= \lVert \vec{x}_{v_x}' - \vec{x}_{q} \rVert_2^2 
    = 
    \lVert \vec{x}_{v_x}' - \vec{x}_{v_y}' + \vec{x}_{v_y}' - \vec{x}_{q} \rVert_2^2 \\
    & = 
    \lVert \vec{x}_{v_y}' - \vec{x}_{v_x}' \rVert^2_2 + \lVert \vec{x}_{v_y}' - \vec{x}_q \rVert^2_2 - 2\langle \vec{x}_{v_y}' - \vec{x}_{v_x}', \vec{x}_{v_y}' - \vec{x}_q\rangle\\
\end{aligned}
\end{equation}

Thus, the distance comparison between $\delta(\vec{x}_{v_x}',\vec{x}_{q})$ and $\delta(\vec{x}_{v_y}',\vec{x}_{q})$ can be calculated as follows.
\begin{equation}
\label{eq:cos_theta}
\small
\begin{aligned}
    \lVert \vec{x}&_{v_x}' - \vec{x}_{q} \rVert_2^2 - \lVert \vec{x}_{v_y}' - \vec{x}_{q} \rVert_2^2  \\
    &= \langle \vec{x}_{v_y}' - \vec{x}_{v_x}', \vec{x}_{v_y}' - \vec{x}_{v_x}' \rangle - 2\langle \vec{x}_{v_y}' - \vec{x}_{v_x}', \vec{x}_{v_y}' - \vec{x}_q\rangle \\
    &=\langle \vec{x}_{v_y}' - \vec{x}_{v_x}', 2\vec{x}_q - \vec{x}_{v_x}' - \vec{x}_{v_y}'  \rangle \\
    &=2\lVert \vec{x}_{v_y}' - \vec{x}_{v_x}' \rVert_2\lVert \vec{x}_q - \frac{\vec{x}_{v_x}' + \vec{x}_{v_y}'}{2} \rVert_2 \cos \theta
\end{aligned}
\end{equation}

According to Eq. \ref{eq:cos_theta}, the comparison between $\delta(\vec{x}_{v_x}',\vec{x}_{q})$ and $\delta(\vec{x}_{v_y}',\vec{x}_{q})$ includes three terms: $\lVert \vec{x}_{v_y}' - \vec{x}_{v_x}' \rVert_2$ indicates the distance between two neighbors, $\lVert \vec{x}_q - \frac{\vec{x}_{v_x}' + \vec{x}_{v_y}'}{2} \rVert_2$ represents the distances between neighbors to the query, and $\cos\theta$ is the angle between the vectors $\vec{x}_{v_x}'-\vec{x}_{v_y}'$ and $\vec{x}_q - \frac{\vec{x}_{v_x}' + \vec{x}_{v_y}'}{2}$.
\end{proof}

\setlength{\textfloatsep}{0.1cm}
\begin{table}
  \setlength{\abovecaptionskip}{0.1cm}
  \centering
  \captionsetup{labelfont=bf}
  \caption{\textbf{Recall@10 when considering different features}}
  \label{tab:motivation}
   \renewcommand\arraystretch{1.6}
  \scalebox{0.9}{
  \begin{tabular}{c||c|c|c|c}
    \hline
    \textbf{Features $\downarrow$}         & \textbf{Sift}   & \textbf{Deep} & \textbf{Ukbench} & \textbf{Gist} \\
    \hline
    \hline
    ranking w/ neighbor \& routing         & 0.700                & 0.710                & 0.790  & 0.732 \\
    \hline
    ranking by Eq. \ref{eq:cos_theta}             & 0.950                & 0.978               & 0.987 & 0.892 \\
    \hline
\end{tabular}
}
\end{table}

%\vspace{-0.3cm}
Table \ref{tab:motivation} shows a comparative experiment on ANNS's effectiveness using different terms in Eq. \ref{eq:cos_theta} for ranking global candidates during the beam search: ranking with the first two terms (1st row) and ranking with three terms (2nd row). The first two terms contribute a lot to the recall, while additional consideration of the third term provides extra improvement. Although the third term is difficult to be mathematically expressed, it is clear that $\cos \theta$ is related to the first two terms. Motivated by this, we propose that optimizing the PQ for graph-based ANNS requires considering not only the neighborhood features but also the PG routing features. 

%Building upon this idea, we present RPQ framework in \S \ref{sec:framework}.

% \vspace{1ex}
%\vspace{-0.1cm}
\subsection{RPQ Framework} \label{sec:framework}
Figure \ref{fig:2} illustrates the pipeline of RPQ that contains three modules: \textit{differentiable quantizer}, \textit{feature extractor}, and \textit{multi-feature joint training}. The differentiable quantizer plays an important role of the entire RPQ, which enables the back-propagation of end-to-end learning. It converts original vectors into compact codes. All compact codes are taken as input of feature extractor to get representative neighborhood and routing features for a specific PG. Finally, the multi-feature joint training continuously optimize the differentiable quantizer using feature-aware losses via back-propagation. 
%We briefly introduce all modules below and \textcolor{blue}{show their details in \S \ref{sec:model}-\S \ref{sec:training}}.

\vspace{0.1cm}
\noindent\textbf{Differentiable quantizer (\S \ref{sec:loss}).} The differentiable quantizer consists of two steps: (1) \textit{adaptive vector decomposition} and (2) \textit{differentiable quantization}. Given an original $D$-dimensional vector $\vec{x}_i$, it first automatically decomposes it into $M$ sub-vectors $\{\vec{x}_i^1,\cdots,\vec{x}_i^M\}$, of which every sub-vector has $D/M$ dimensions with balanced valuable information. Then, it converts each $\vec{x}_i^j$ to a compact code $\mathcal{Q}(\vec{x}_i^j)$ in a continuous space via a differentiable transformation. 
%Given the compact code, we take them as input of feature extractor for obtaining neighborhood and routing features.

% In RPQ, we propose a learnable differentiable quantizer, which embeds neighborhood and routing features into both the codebook and compact codes, thus generating high-quality quantized vectors facilitating graph-based ANNS's effectiveness.

\vspace{0.1cm}
\noindent\textbf{Feature extractor (\S \ref{sec:model}).}  Feature extractor aims to extract the representative neighborhood and routing features for a specific PG, using the aforementioned quantized results.

\vspace{0.1cm}
\noindent\underline{Neighborhood features.} Given a PG $G$ = $(V,E)$ and a node $u\in V$, we use $\vec{x}_u$ with a node specifier to indicate its corresponding original vector in the vector dataset $\mathcal{X}$ because there is a bijection from $\mathcal{X}$ to $V$. We expect that for each neighbor $v$ of $u$ in $G$, i.e., $\forall v\in N(u)$, their quantized vectors $\vec{x}_u'$ and $\vec{x}_v'$ are closer if their original vectors $\vec{x}_u$ and $\vec{x}_v$ are closer in $\mathcal{X}$, and vice versa. To achieve this, RPQ employs contrastive learning based on a set of triplets to embed the neighborhood relationship into the differentiable quantizer. We define the positive and negative sample of a triplet as follows.
%Given a PG $G$ and a node $u\in V$, we expect that for each neighbor $v$ of $u$ in $G$, i.e., $\forall v\in N(u)$, their quantized vectors $\vec{x}_u'$ and $\vec{x}_v'$ are closer if their original vectors $\vec{x}_u$ and $\vec{x}_v$ are closer in the original vector dataset $\mathcal{X}$, and vice versa. To achieve this, RPQ employs contrastive learning based on a set of triplets to embed the neighborhood features into learned PQ. We define the positive and negative sample of a triplet as follows.
%reconstructed by compact codes and codebook

\vspace{0.1cm}
\begin{myDef}
\label{def:positive_sample}
\textbf{Positive sample.} Given a PG $G(V,E)$ and a vertex $v\in V$, we define a positive sample of $v$'s quantized vector $\vec{x}_v'$ as $\vec{x}_{v_+}'$, where $v_+$ is one of the $k_{\rm pos}$ nearest neighbors from $v$'s neighborhood. We use $k_{\rm pos}$ to control the scope from where the positive sample comes.
\end{myDef}

\vspace{0.1cm}
\begin{myDef}
\label{def:negative_sample}
\textbf{Negative sample.} Given a PG $G(V,E)$ and a vertex $v\in V$. We define a negative sample of $v$'s quantized vector $\vec{x}_v'$ as $\vec{x}_{v_-}'$, where $v_-$ is one of the $k_{\rm neg}$ nearest neighbors (besides the former $k_{\rm pos}$ neighbors) from the neighborhood of $v$. Here, we use a parameter $k_{\rm neg}$ to control the scope from where the negative sample comes.
\end{myDef}

Given a set of triples $\{\langle \vec{x}_{v_+}', \vec{x}_v', \vec{x}_{v_-}'\rangle\}$, RPQ aims to make those closely connected neighbors $\langle \vec{x}_{v_+}', \vec{x}_v'\rangle$ more closer and other neighbors $\langle \vec{x}_v', \vec{x}_{v_-}'\rangle$ as far away as possible.
%Thus, given a vector $\vect{x}_v$ for $\forall v\in V$ as an anchor, we have a triplet as $\langle \vec{x}_{v_+}, \vec{x}_v, \rangle \vec{x}_{v_-}$

%Here, we use a parameter $k_{\rm pos}$ ($k_{\rm neg}$) to control the scope from where the positive (negative) sample comes from. In \S xx, we 

\vspace{0.1cm}
\noindent\underline{Routing features.} Given a query vector $\vec{x}_q$ and a PG $G$, the routing is performed as a beam search over $G$ that starts from an entry vertex $v_e$ to the $k$ vertices whose quantized vectors are the closest to $\vec{x}_q$. We consider the decision-making process at each visited vertex during the entire routing as the routing features. Intuitively, the more good decisions made during the routing, the more likely the search would locate in the neighborhood of $\vec{x}_q$ and return the nearest neighbors of $\vec{x}_q$. We define the routing features as follows.
%Since the routing decision is made based on \textcolor{blue}{the ADC distances between all candidates and $\vec{x}_q$}, 

%Given a query vertex $\vec{x}_q$ and an entry vertex $\vec{x}'_e$, the routing path from $\vec{x}'_e$ to $\vec{x}'_q$ is denoted by $\vec{x}'_e\rightsquigarrow \vec{x}'_q$. Comparing with recording the entire routing path as routing features, we prefer to consider the decision-making process at each vertex $\vec{x}'_v\in \vec{x}_e\rightsquigarrow \vec{x}'_q$ as the routing features. Intuitively, the more good decisions made during the routing procedure, the more likely the search would locate in the neighborhood of $\vec{x}'_q$ and return the nearest neighbors of $\vec{x}'_q$. Since the routing decision is made based on the distances between all candidates and $\vec{x}'_q$, we define the routing features as follows.

\vspace{0.1cm}
\begin{myDef}
\label{def:routing_feature}
\textbf{Routing features.} Given a PG $G(V,E)$, a query vector $\vec{x}_q$, and a real routing process starting from an entry vertex $v_e$. We define the routing features for $\vec{x}_q$ as a set $\mathcal{B}(\vec{x}_q)=\{b_{1},\cdots,b_{L}\}$. Here, each $b_i=\{\mathcal{Q}((\vec{x}_{v_1}),\cdots,\mathcal{Q}(\vec{x}_{v_h})\}$ records the ranked compact codes (in ascending order of $\delta(\vec{x}_{v_j}',\vec{x}_q)$) of $h$ candidate vertices for the next-hop decision at the $i$-th step of beam search, and $L \in \mathbb{N}$ is the number of decisions have made during the routing of beam search.
\end{myDef}

%Given a PG $G(V,E)$, a \textcolor{blue}{query vector $\vec{x}_q$}, and a real routing path $\vec{x}'_e\rightsquigarrow \vec{x}'_q$ from an entry vertex $\vec{x}'_e$, we define the routing features as a set $\mathcal{B}(\vec{x}'_q)=\{b_{1},\cdots,b_{L}\}$. Here, each $b=\{\vec{x}'_{v_1},\cdots,\vec{x}'_{v_h}\}$ denotes $h$ ranked candidates (in ascending order of distance to $\vec{x}'_q$) at a vertex $\vec{x}'_v\in \vec{x}'_e\rightsquigarrow \vec{x}'_q$, and $L \in \mathbb{N}$ is the number of decisions have made during routing procedure.

\vspace{0.1cm}
\noindent\textbf{Multi-feature joint training module (\S \ref{sec:training}).} Finally, we take above neighborhood and routing features as training data to optimize the differentiable quantizer with two feature-aware losses. In this way, the neighborhood and routing features would be embedded into the learned quantizer, thus facilitating the graph-based ANNS.

%thus making the quantizer more adaptive to graph-based ANNS methods.

 \section{Differentiable quantizer} 
\label{sec:model}
%We discuss two steps of our differentiable quantizer below.

\vspace{0.1cm}
\noindent\textbf{Adaptive vector decomposition.} Existing PQ methods usually apply vertical division to divide a $D$-dimensional vector into $M$ sub-vectors, of which the first $D/M$ dimensions belongs to the first sub-vector and the $[D/M+1,2D/M]$ dimensions belong to the second sub-vector, etc. In this way, it causes a critical issue, that is: the dimensions with valuable intrinsic features would unbalancedly locate in each sub-vector \cite{ge2013optimized}, resulting in meaningless quantized sub-vectors from those sub-vectors with less intrinsic features. 

To handle this, we present an adaptive vector decomposer, to automatically determine which dimensions belong to each sub-vector by applying a square orthonormal matrix $R$ for all vectors to make the valuable dimensions uniformly distributed among all sub-vectors \cite{ge2013optimized, norouzi2013cartesian}. Specifically, our adaptive vector decomposition involves following steps. (1) We introduce a skew-symmetric matrix $A \in \mathbb{R}^{D \times D}$ as a group of learnable parameters. (2) We then use exponential algorithm to initialize the square orthonormal matrix $R$ based on $A$, denoted by $R= exp(A)$, where $exp(\cdot)$ indicates the matrix exponential. The orthogonality follows from $exp(A)^T= exp(A^T) = exp(-A) = exp(A)^{-1}$. (3) Given a vector $\vec{x}_v\in \mathcal{X}$, we rotation it as $R\vec{x}_v$, thus balancing the informativeness across the entire vector. (4) We divide the rotated vector $R\vec{x}_v$ into $M$ sub-vectors $\{R\vec{x}_v^1, \cdots, R\vec{x}_v^M\}$.
%, of which every sub-vector has $D/M$-dimensions. 
%Make the valuable dimensions equally dispersed among all sub-vectors

We take the decomposed sub-vectors as input of differentiable quantization and update the learnable parameters $A$ by multi-features joint training (\S \ref{sec:training}) using the features obtained in \S \ref{sec:loss}, thus continuously optimizing $A$ to generate a good $R$ for better vector decomposition. {\cite{ge2013optimized} presents a covariance matrix to measure the value of each dimension in a vector. We use it to visualize the original and optimized distribution of valuable dimensions before and after adaptive vector decomposition in Figure \ref{fig:valuedist} as a case study. Take an vector with 128 dimensions from Sift1M as an example (8 rows of sub-vectors with 16 dimensions for each are provided on the left part), the right part shows the sub-vectors after 100 iterations of optimization on $A$. It is evident that the valuable dimensions are uniformly distributed among all sub-vectors.

%where two vectors with 128 dimensions from Sift 1M and (b) is a vector from Deep 1M, and  . Note that, the redder the color, the more important the dimension. It is evident that after 100 iterations of optimization on $A$, the valuable dimensions are uniformly distributed among all sub-vectors.}

\begin{figure}
%\vspace{-0.2cm}
\includegraphics[width=90mm]{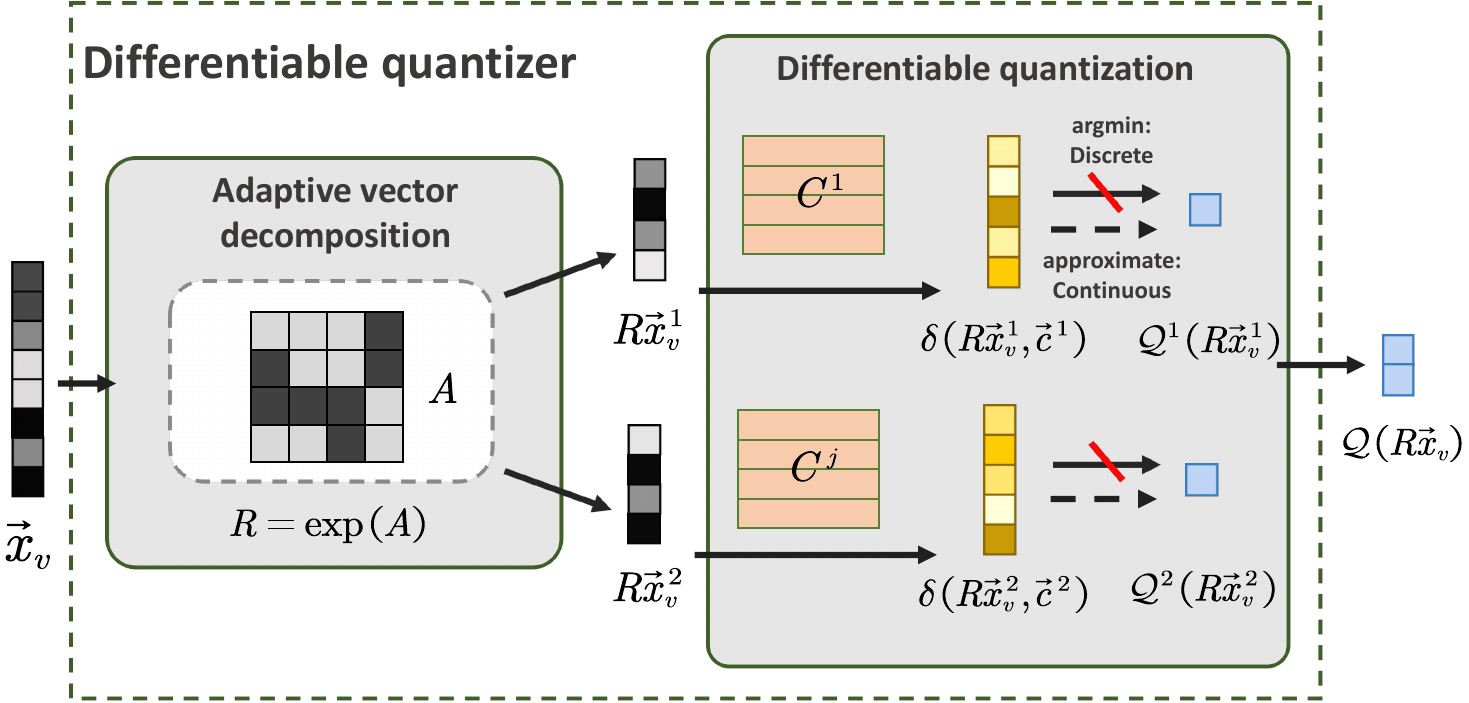}
\vspace{-0.6cm}
\caption{\textbf{Differentiable quantizer}}\label{fig:3}
\vspace{-0.1cm}
\end{figure}

\vspace{0.1cm}
\noindent\textbf{Differentiable quantization.} Given a set of rotated sub-vectors $\{R\vec{x}_v^1,\dots, R\vec{x}_v^M\}$ of $\vec{x}_v$ and a codebook $\mathcal{C}=\{\mathcal{C}^1,\dots,\mathcal{C}^M\}$ of which each $\mathcal{C}^j$ involves $K$ codewords, we can straightforwardly quantize sub-vectors as follows. First, for each $R\vec{x}_v^j$, we compute its distance to every codeword from the $j$-th codebook $\mathcal{C}^j$. Then, we apply the \emph{argmin} function to select the closest codeword to $R\vec{x}_v^j$ and use its identifier as the compact code of $R\vec{x}_v^j$, denoted by $\mathcal{Q}(R\vec{x}_v^j)$. Given a set of compact codes $\{\mathcal{Q}^1(R\vec{x}_v^1),\dots, \mathcal{Q}^M(R\vec{x}_v^M)\}$, we can easily get their quantized sub-vectors $\{\mathcal{C}^1(\mathcal{Q}^1(R\vec{x}_v^1)),\dots, \mathcal{C}^M(\mathcal{Q}^M(R\vec{x}_v^M))\}$ according to Definition \ref{def:pq}. However, we face a challenge to use this
quantized sub-vectors for training, that is: the \emph{argmin} function is non-differentiable, thus making back-propagation invalided from the compact code to its corresponding sub-vector.

%Since the codeword assignment probability and Gumbel-Softmax are differentable, we can conduct back-propagation s
To handle this problem, we present a differentiable approximation instead of the \textit{argmin} function to encode a sub-vector as compact code. First, we compute a \textit{codeword assignment probability} for each sub-vector, showing the probability of this sub-vector would be encoded by a specific codeword. Then, we use \textit{Gumbel-Softmax}\cite{maddison2016concrete, jang2016categorical} to compute an approximate compact code based on above assignment probabilities. Since the codeword assignment probability and Gumbel-Softmax are differentable, the whole quantization is also differentable.

\vspace{0.1cm}
\noindent\underline{Codeword assignment probability.} Given a rotated sub-vector $R\vec{x}^j$, a codebook $\mathcal{C}^j$, we compute the probability of $R\vec{x}^j$ is encoded by the $k$-th codeword from $\mathcal{C}^j$ as Eq. \ref{eq:7}.
\begin{equation}
    p(\vec{c}_{k}^j|{R\vec{x}}^j)=\frac{\exp(\delta({R\vec{x}}^j, \vec{c}_{k}^{\,j}))}{\sum\limits^{K}_{k=1} \exp(\delta({R\vec{x}}^j, \vec{c}_{k}^{\,j}))}  \label{eq:7}
\end{equation}

%\vspace{0.1cm}
\noindent\underline{Approximate compact code.} By using Eq. \ref{eq:7}, we can get $K$ probabilities $\{p(\vec{c}_{1}^j|{R\vec{x}}^j),\dots,p(\vec{c}_{K}^j|{R\vec{x}}^j)\}$ of a sub-vector $R\vec{x}^j$ to be encoded by $K$ codewords from $\mathcal{C}^j$. Next, we apply Gumbel-Softmax on these probabilities to get an approximate compact code of $R\vec{x}^j$ by Eq. \ref{eq:gumbel-softmax}:
\begin{equation}
\label{eq:gumbel-softmax}
    \mathcal{Q}^j({R\vec{x}}^j)={\rm softmax}\{\log p(\vec{c}_{k}^{\,j}|{R\vec{x}}^j)+z_k,k=1,\dots,K\}\quad ,
\end{equation}
\noindent where $z_k$ is a sample from the standard Gumbel distribution that can be obtained as \emph{$z_k=-\log(-\log {\rm Uniform}(0,1))$}. Since Eq. \ref{eq:7} and Eq. \ref{eq:gumbel-softmax} are differentiable, it is possible to use the loss (discussed in \S \ref{sec:loss}) to update the square orthonormal matrix $R$ via back-propagation, therefore update the skew-symmetric matrix $A$ because of $R=\exp(A)$. The better the $A$ is, the better the sub-vectors are obtained.

\begin{figure}
\centering
%\vspace{-0.2cm}
\includegraphics[width=80mm]{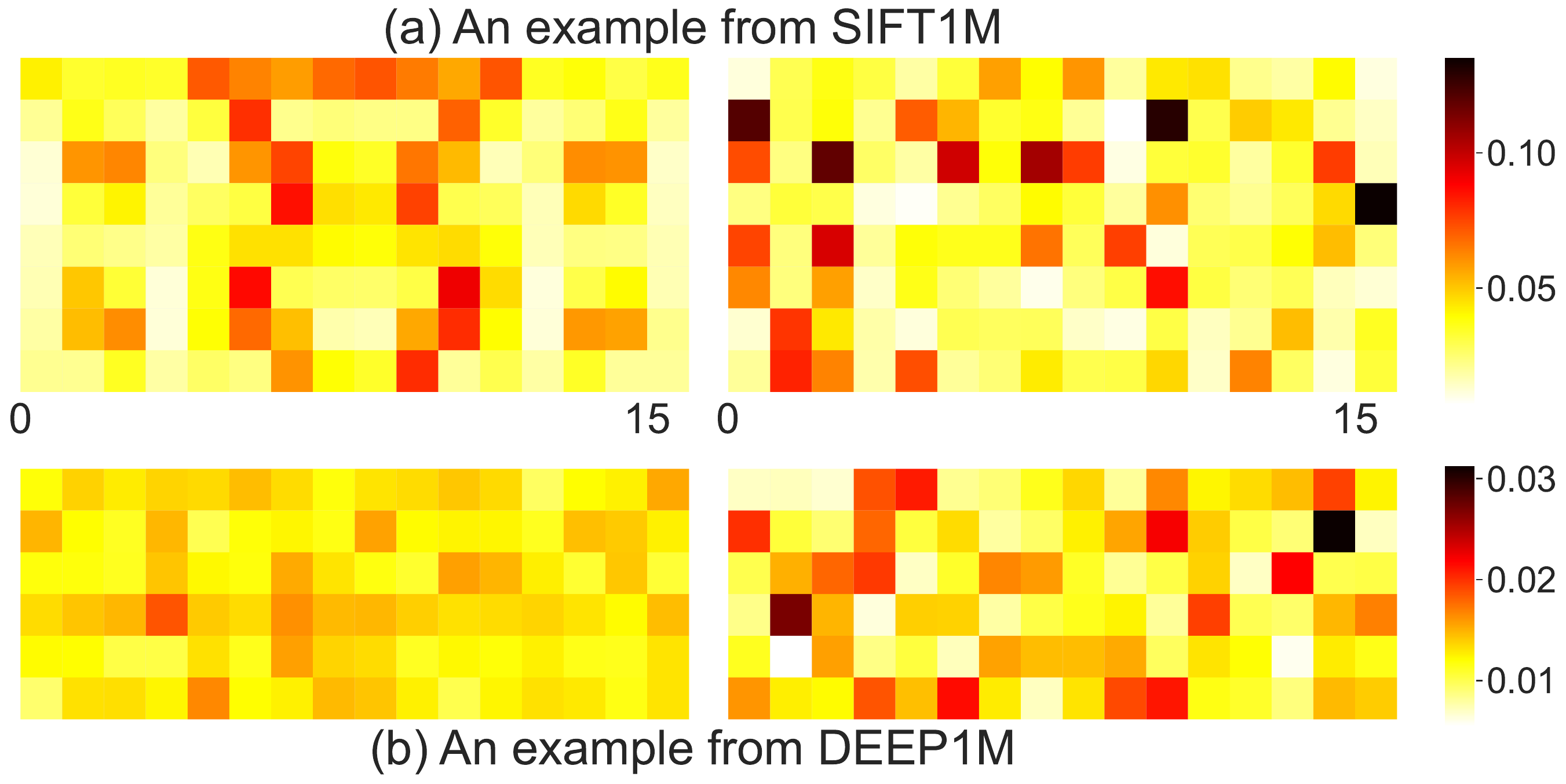}
\vspace{-0.4cm}
\caption{\textbf{A case study of valuable dimensions' distribution (the redder the color, the more the value of a dimension)}}
\label{fig:valuedist}
%\vspace{-0.1cm}
\end{figure}

\vspace{0.2cm}
\section{Sampling-based Feature Extractor}  \label{sec:loss}
In this section, we present a sampling-based method to extract both neighborhood and routing features for end-to-end PQ learning.

\vspace{0.1cm}
\noindent\textbf{Neighborhood features sampling.} Given a vertex $v\in V$ with a quantized vector $\vec{x}_v'$, where $V$ is a vertex set of a PG $G=(V,E)$, a straightforward sampling method is to take all $v$'s 1-hop neighbors $N(v)$ as the population, and conduct a random sampling to collect one positive sample as the quantized vector $\vec{x}_{v_+}'$ of a vertex $v_+$ from $N(v)$ and collect one negative sample as the quantized vector $\vec{x}_{v_-}'$ of a vertex $v_-$ from the remaining vertices $V\setminus N(v)$. This method is easy to implement, however, it suffers from one critical issue, that is: in most of the popular PGs, e.g., HNSW \cite{malkov2018efficient}, NSG \cite{fu2017fast}, Vamana \cite{jayaram2019diskann}, a vertex's neighbors may not be its top nearest vertices. This is because a PG often leverage some secondary nearest vertices serving as \textit{highways} or \textit{shortcuts} to connect remote vertices, thus enhancing the ability of searching a query vertex that is far away from the entry \cite{malkov2014approximate, prokhorenkova2020graph, malkov2018efficient}. This inspires us to present an $n$-propagation sampling method to collect positive/negative samples from a $n$-hop neighborhood of a vertex $v$.

\begin{algorithm}[t]
%\DontPrintSemicolon
    \SetAlgoLined
    \KwIn {vertex $v \in V$, quantizer $\mathcal{Q}$, codebook $\mathcal{C}$, hop number $n$, threshold of sampling $k_{\rm pos}$ and $k_{\rm neg}$}
    \KwOut {triplet $\langle \vec{x}_{v_+}',\vec{x}_v',\vec{x}_{v_-}' \rangle$}

    $N_{n}(v) \leftarrow \emptyset, S \leftarrow N(v), {\rm {\sf Visit}} \leftarrow \emptyset$ \;
    \For {$1$ to $n$} {
        \For{$\forall v_i \in S$}{
            $N_{n}(v) \leftarrow N_{n}(v) \cup v_i, {\rm {\sf Visit}} \leftarrow {\rm {\sf Visit}} \cup v_i$ \;
            \For{$\forall v_j \in N(v_i)$} {
                $N_{n}(v) \leftarrow N_{n}(v) \cup v_j$ \;
            }
        }
        $S \leftarrow N_n(v) \setminus {\sf Visit}$ \;
    }
    sort $N_n(v)$ in ascending order of the distance to $\vec{x}_v$ \;
    $N_n(v).$resize$(k_{\rm pos} + k_{\rm neg})$ \;
    $N_{\rm pos}(v)\leftarrow \emptyset, N_{\rm neg}(v)\leftarrow \emptyset$ \;
    \For {$i = 1$ to $k_{\rm pos}$} {
        $N_{\rm pos}(v) \leftarrow N_{\rm pos}(v) \cup N_n(v)[i]$ \;
    }
    $N_{\rm neg}(v) \leftarrow N_n(v) \setminus N_{\rm pos}(v)$ \; 
    $v_+\leftarrow$ a random sample from $N_{\rm pos}(v)$ \;
    $v_-\leftarrow$ a random sample from $N_{\rm neg}(v)$) \;
    $\vec{x}_{v_+}' = \mathcal{C}(\mathcal{Q}(\vec{x}_{v_{+}})), \vec{x}_{v_-}' = \mathcal{C}(\mathcal{Q}(\vec{x}_{v_{-}}))$ \;
    \Return $\langle \vec{x}_{v_+}',\vec{x}_{v},\vec{x}_{v_-}' \rangle$
    \caption{$n$-propagation sampling}  \label{algo:1}
\end{algorithm}

Alg. \ref{algo:1} shows the procedure of $n$-propagation sampling. First, we collect a vertex $v$'s $n$-hop neighbors $N_{n}(v)$ as population (lines 2-10), where $S$ records vertices for propagation and {\sf Visit} is used to avoid duplicated visiting. Then, we collect positive and negative samples from $N_{n}(v)$ as follows.

\vspace{0.1cm}
\noindent\underline{Positive sampling.} Given the population $N_{n}(v)$, we introduce a parameter $k_{\rm pos}\in [1, |N_n(v)|)$ to indicate the top-$k_{\rm pos}$ nearest vertices that form the sampling scope of positive samples (lines 14-16). The larger the $k_{\rm pos}$, the more neighbors considered as candidates of positive sample, thus resulting in oversampling where some secondary nearest vertices would be sampled. In contrast, the smaller the $k_{\rm pos}$, the less neighbors considered as candidates of positive sample. This would result in undersampling where insufficient neighborhood features are considered. In our experimental study (\S \ref{sec:experiments}), we show $k_{\rm pos}$'s effect on learned PQ's effectiveness.

\vspace{0.1cm}
\noindent\underline{Negative sampling.} Given the population $N_{n}(v)$, we introduce a parameter $k_{\rm neg}\in [1, |N_n(v)|-k_{\rm pos}]$ to indicate $k_{\rm neg}$ secondary nearest vertices that form the sampling scope of negative samples (line 17). If $k_{\rm neg}=|N_n(n)|-k_{\rm pos}$, then it means we collect a negative sample from all remaining vertices besides top-$k_{\rm pos}$ nearest vertices. In fact, among these $|N_n(n)|-k_{\rm pos}$ secondary nearest vertices, some vertices are closer to $v$ than others and they usually are called as \textit{hard samples} and often more valuable for learning than others \cite{Cohan2020,xu2022academic,Wang2023}. If we can distinguish the hard samples from the top-$k_{\rm neg}$ nearest vertices, then the far away vertices from $v$ probably be distinguished easily. We use $k_{\rm neg}$ to control the scope of negative sample comes from. The smaller the $k_{\rm neg}$, the more hard samples would be considered. The larger the $k_{\rm neg}$, the more simple samples would be considered. It is widely-recognized that a balance between hard and simple samples is important to learning \cite{zhan2021optimizing}, therefore in \S \ref{sec:experiments}, we show $k_{\rm neg}$'s effect on learned PQ's effectiveness.

\vspace{0.1cm}
\noindent\textbf{Routing features sampling.} In \S \ref{sec:framework}, we define the routing features as all ranked candidates considered for next-hop selection during the entire beam search. Alg. \ref{algo:2} shows the procedure of the routing features sampling. We first collect a set of vectors from the vector dataset as the query samples, denoted by $\{\vec{x}_{q_1},\cdots,\vec{x}_{q_n}\}$ (line 1). Then, for each query vector $\vec{x}_q$, we start from an entry vertex $v_e$ (i.e., the global candidate set $b$ is initialized as $\{\mathcal{Q}(\vec{x}_{v_e})\}$) to perform beam search (lines 4-16). Specifically, at each step, we select the next-hop as the closest unvisited vertex $v^*$ to $\vec{x}_q$ by exploring its neighbors and updating $b$ with these neighbors' compact codes (lines 6-10). Next, we rank all global candidates in ascending order of their distances to $\vec{x}_q$ using ADC (line 11) and maintain the global candidate set with exactly $h$ elements (lines 12-14). Then, we add the global candidates $b$ at each step into $\mathcal{B}_q$ and repeat above until all vertices in $b$ have been visited (line 15). Finally, we add each $B_q$ for each $\vec{x}_q$ into $B_{\rm query}$ and return it as the routing features (line 17 \& 19). In a nutshell, suppose an entire beam search for a query $\vec{x}_q$ involves $l$ next-hop selections, then we have $\mathcal{B}_q=\{b_{1},\cdots,b_{l}\}$, of which each $b_i=\{\mathcal{Q}(\vec{x}_{v_1}),\cdots,\mathcal{Q}(\vec{x}_{v_h})\}$ denotes $h$ ranked candidates for the $i$-th next-hop selection.

%we rank all global candidates from $b$ in ascending order of their distances to $\vec{x}_q$, i.e., $\delta(\vec{x}_{v_n}',\vec{x}_q,)$ using ADC (line 12), and we select the top-1 unvisited closest candidate to $\vec{x}_q$ as the next-hop for search expansion (line 4) until no more candidates for expansion (line x)}. In a nutshell, suppose an entire beam search involves $l$ next-hop selections and each selection requires to consider $h$ candidates, then we record a set $\mathcal{B}(\vec{x}_q')=\{b_{1},\cdots,b_{l}\}$ as the routing features, of which each $b_i=\{\vec{x}_{v_1}',\cdots,\vec{x}_{v_h}'\}$ denotes $h$ ranked candidates for the $i$-th next-hop selection.

%To achieve such routing features, we do the following: First, we randomly generate a set of queries $\{\vec{x}_{q_1},\cdots,\vec{x}_{q_n}\}$ and \textcolor{blue}{perform beam search for each query using ADC for quantized vectors.} Alg. \ref{algo:2} shows the procedure of the routing features sampling. Specifically, we start from the entry $\vec{x}_e'$ and initialization candidate with $\vec{x}_e'$ (lines 2). Then, we select the top-1 closest candidate to $\vec{x}_q'$ as the next-hop for search expansion until no more candidates for expansion (lines 3-16). In a nutshell, suppose a routing path involves $l$ next-hop decisions and each decision requires to consider $h$ candidates, then we record the set $\mathcal{B}(\vec{x}_q')=\{b_{1},\cdots,b_{l}\}$ as the routing features, of which $b=\{\vec{x}_{v_1}',\cdots,\vec{x}_{v_h}'\}$ denotes $h$ ranked candidates (in ascending order of distance to $\vec{x}_q'$).

\begin{algorithm}[t]
%\DontPrintSemicolon
    \SetAlgoLined
    \KwIn {quantizer $\mathcal{Q}$, codebook $\mathcal{C}$, entry vertex $v_e$, vector dataset $\mathcal{X}$, $h$ global candidates}
    \KwOut {candidates set $\mathcal{B}_{\rm query}$ for $\mathcal{X}_{\rm query}$}
    $\mathcal{X}_{\rm query}\leftarrow$ a set of query samples collected from $\mathcal{X}$ \;
    $\mathcal{B}_{\rm query} \leftarrow \emptyset $ \;

    \For {$\forall \vec{x}_{q} \in \mathcal{X}_{\rm query}$} {
        $\mathcal{B}_{q} \leftarrow \emptyset, b \leftarrow \{\mathcal{Q}(\vec{x}_{v_e})\}, {\sf Visit} \leftarrow \emptyset$ \;

        \While{$\exists v \in b$ {\rm \textbf{and}} $v \notin {\sf Visit}$} {
            $v^*\leftarrow$ the closest vertex $\in b$ to $\vec{x}_q$ \textbf{and} $\notin {\sf Visit}$ \;
            ${\sf Visit} \leftarrow {\sf Visit} \cup v^*$ \;
            \For{$\forall v_j \in N(v^*)$ {\rm \textbf{and}} $v_j \notin {\sf Visit}$} {
                    $b \leftarrow b \cup \mathcal{Q}(\vec{x}_{v_j})$ \;
            }
            sort $b$ in ascending order of the distance to \textbf{$\vec{x}_q$} \;
            \If{$b$.size() $> h$} {
                $b$.resize($h$) \;
            }
            $\mathcal{B}_{q} \leftarrow \mathcal{B}_{q} \cup b$ \;
        }
        $\mathcal{B}_{\rm query} \leftarrow \mathcal{B}_{\rm query} \cup \mathcal{B}_{q}$
    }
    \Return $\mathcal{B}_{\rm query}$
    \caption{routing features sampling}  \label{algo:2}
\end{algorithm}

%\vspace{-0.1cm}
\section{Multi-feature joint training module} \label{sec:training}
%\vspace{-0.1cm}
We consider both the neighborhood and routing features to design two feature-aware losses and combine them as a \textit{multi-feature joint loss} to optimize the differentiable quantizer. The feature-aware losses includes two parts: the neighborhood feature loss and the routing feature loss. For the neighborhood features in forms of triplets $\{\langle \vec{x}'_{v_+},\vec{x}'_v,\vec{x}'_{v_-} \rangle\}$, we employ contrastive learning to minize the triplet loss in the quantization space. Intuitively, we want $\vec{x}_v'$ to be closer to it's positive sample $\vec{x}_{v+}'$ than to the negative sample $\vec{x}_{v-}'$. The triplet loss is provided in Eq. \ref{eq:nein}, where $\sigma$ is the margin hyperparameter.
\begin{equation}
    \mathcal{L}_{\rm neighborhood} = max(0, \sigma + \delta(\vec{x}_v', \vec{x}_{v+}') - \delta(\vec{x}_v', \vec{x}_{v-}'))   \label{eq:nein}
\end{equation}
%and optimizing the differentiable quantizer represented in \S \ref{sec:loss}. 
%We expect that using the quantized vectors for graph-based ANNS can achieve a comparable effectiveness with that using the original vectors. 

For routing features, suppose $\mathcal{Q}(\vec{x}_{v^*})$ is the compact code of next-hop selected vertex $v^*$ from the given candidates $b_i$ at the $i$-th next-hop selection of the beam search for query $\vec{x}_{q}$ (line 6 of Alg. \ref{algo:2}). Then, the goal of learning is to maximize a conditional probability of choosing $\mathcal{Q}(\vec{x}_{v^*})$ from $b_i$ (Eq. \ref{eq:nexthop}).
\begin{equation}
\label{eq:nexthop}
    P(\mathcal{Q}(\vec{x}_{v^*}) | \vec{x}_{q}, b_i) = \frac{e^{(\delta(\mathcal{C}(\mathcal{Q}(\vec{x}_{v^*})), \vec{x}_{q})/\tau}}{\sum\limits_{\mathcal{Q}(\vec{x}_{j}) \in b_i} e^{(\delta(\mathcal{C}(\mathcal{Q}(\vec{x}_{j})),\vec{x}_{q}) / \tau}}  
\end{equation}

To achieve this, we establish the routing feature loss by maximizing the log-likelihood of optimal next-hop selection.
%Hence, RPQ maximizes the log-likelihood of optimal decisions as follows:
\begin{equation}
\begin{aligned}
    \mathcal{L}_{\rm routing} = - \sum\limits_{\mathcal{B}_{q} \in \mathcal{B}_{\rm query}} \sum\limits_{b_i \in \mathcal{B}_{q}} \log P(\mathcal{Q}(\vec{x}_{v^*}) | \vec{x}_{q}, b_i) 
\end{aligned}
\end{equation}

Finally, we use the sum of above two losses with a learnable coefficient $\alpha$ as the multi-feature joint loss.
\begin{equation}
    \mathcal{L} = \mathcal{L}_{\rm routing} + \alpha \mathcal{L}_{\rm neighborhood}
\end{equation}

We aim to minimize the joint loss $\mathcal{L}$ using mini-batch gradient descent with the \textit{Adam optimizer} \cite{kingma2014adam}. We use one cycle learning rate schedule for faster model convergence and the hyperparameters are: LR $=$ 1e-3, decay rate $=$ 0.2. 

\vspace{-0.1cm}
\section{Integration of Learned PQ with Existing Graph-based ANNS} \label{sec:search}
%In this section, we show how to integrate our routing-guided learned PQ with existing graph-based ANNS. 

According to whether the external drive is involved, the application of using PQ to optimize memory usage is generally divided into two categories: (1) PQ-integrated graph-based ANNS for hybrid scenario \cite{jayaram2019diskann, ren2020hm}, such as DiskANN \cite{jayaram2019diskann} and its variants Filter-DiskANN \cite{gollapudi2023filtered}, OOD-DiskANN \cite{jaiswal2022ood}, and Fresh-DiskANN \cite{singh2021freshdiskann}. They only record the small volume of compact codes and codebook in memory, while the large volume of original vectors and PG index are retained in SSD. The search requires the participation of both the PQ distance and original vector distance. (2) PQ-integrated graph-based ANNS for in-memory scenario \cite{douze2018link, faissweb}. In this case, the usual practice is to replace the original vectors with a smaller amount of compact codes and codebook, and store them in memory together with the PG. Different from the hybrid scenario, the search in this case only relies on the PQ distance. Both scenarios assume that the memory is too small to load all of the original vectors. For the former, it is usually less efficient than the latter due to the extra SSD I/Os, but the effectiveness is largely improved by using the original distance. Both two are individually suitable for distinct application demands, taking into account the balance between recall and delay\cite{simhadri2022results, guo2022manu}.

%For the latter, it is more efficient as all computations are performed in memory but the effectiveness is affected because only the PQ distance is used.

%\textcolor{blue}{We next discuss the integration our RPQ with graph-based ANNS for both hybrid and in-memory scenarios.}

\vspace{0.1cm}
\noindent \underline{Integration of RPQ for hybrid scenario}. We take DiskANN as an example, when integrating RPQ with DiskANN, this process is made straightforward through the framework inherent to DiskANN. First, we replace DiskANN's PQ with the differential quantizer of our RPQ. During the query processing phase, given a query $\vec{x}_q$, we first divide it into sub-vectors using the orthonormal matrix $R$, then apply ADC to pre-compute the distances between each sub-vector $R\vec{x}_q^{\, i}$ and all sub-codewords, and maintain these distances in a lookup table. Finally, the beam search is gradually performed in the same way as DiskANN: It first obtains a set of candidates for next-hop selection at each visited vertex by quickly checking the lookup table instead of computing the full distance using original vectors. Then, it selects the next-hop vertex using the original vectors resident in SSD via distance reranking.

%We take DiskANN as an example, when integrating RPQ with DiskANN, this process is made straightforward through the framework inherent to DiskANN. \textcolor{blue}{First, we can seamlessly replace DiskANN's PQ method with the differential quantizer of our RPQ. Additionally, unlike DiskANN, an orthonormal matrix $R$ needs to be stored. The size of $R$ is quite small (e.g., 64 KB for BigANN dataset), with a space complexity of $O(D^2)$. During the query processing phase, the query $\vec{x}_q$ is initially transformed into $R\vec{x}_q$. Subsequently, we apply ADC to compute the PQ distance of each quantized vector to $R\vec{x}_q$ in the same way as DiskANN.}
%In this paper, we integrate our RPQ with DiskANN. 
% \subsection{Integrated with In-Memory PGs} \label{sec:base retrieval}

\vspace{0.1cm}
\noindent \underline{Integration of RPQ for in-memory scenario}. The integration process starts by replacing the original vectors with the differentiable quantizer of our RPQ. For each query vector, we also require to first initialize the lookup datable by pre-computing PQ distances between each sub-query vector and $K$ sub-codewords via ADC. Then, the routing process is gradually performed to select the next-hop at each visited vertex by checking the lookup table instead of computing the full distance using original vectors. Different from DiskANN, PQ-integrated graph-based ANNS for in-memory scenario doesn't have the step of distance reranking and it selects the next-hop only based on the PQ distance using lookup datable.

%Since PQ-integrated graph-based ANNS often select HNSW and NSW as the underling PG, 
%\textcolor{blue}{The integration process starts by replacing the original vectors with the differentiable quantizer of our RPQ. During the query processing phase, we apply ADC to calculate the PQ distance. The query $\vec{x}_q$ is initially transformed into $R\vec{x}_q$ using the orthonormal matrix. Subsequently, $R\vec{x}_q$ is divided into several sub-vectors, and we pre-compute the distances between each sub-vector $\vec{x}_q^j$ and the $K$ sub-codewords $\vec{c}^{j}_k$. These distances are maintained in a lookup table for efficient retrieval. Finally, the routing process is gradually performed to select the next-hop at each visited vertex by checking the lookup table instead of computing the distance using original vectors.}

%In-memory PGs currently have various algorithms with different topological structures \cite{wang2021comprehensive, fu2017fast, malkov2018efficient}. However, it is worth noting that our method can effectively adapt to different PGs, and we employ the same integration method for various in-memory PG algorithms. 

\section{Experiments} \label{sec:experiments}
%\href{https://github.com/Lsyhprum/BREWESS.git}{https://github.com/Lsyhprum/BREWESS.git}
We present experiment results of our RPQ on five real-world datasets. The code and datasets have been made available at \cite{BREWESS}. Our evaluation seeks to address the following questions:

\vspace{0.1cm}

\noindent \textbf{Q1:} How do RPQ and other PQ methods perform in terms of effectiveness and efficiency for graph-based ANNS? (\textbf{\S \ref{sec:efficiency}})

\noindent \textbf{Q2}: How do different strategies contribute to RPQ? (\textbf{\S \ref{sec:ablation}})

\noindent  \textbf{Q3}: How do parameters affect RPQ's performance? (\textbf{\S \ref{sec:parameter}})

\noindent  \textbf{Q4}: How is the scalability of RPQ on the data scale? (\textbf{\S \ref{sec:scalability}})

\subsection{Experimental Setting}

\noindent \textbf{Datasets}. Table \ref{tab:dataset} shows the characteristics of widely used real-world datasets for performance evaluation. %Each dataset consists of a base set, query set, and ground-truth set.
\begin{itemize}
\item \textbf{BigANN}\cite{baranchuk2021benchmarks} includes descriptors extracted from an image dataset. We employed slices ranging in size from 1M to 1B, (1M, 10M, 100M, 1B) for scalability evaluation.
\item \textbf{Deep}\cite{YandexBenchmark} includes the projected and normalized outputs from the last fully-connected layer of the GoogLeNet that was pretrained on the Imagenet classification task. Similar to BigANN, we employed four slices for evaluation too.
%\item \textcolor{blue}{\textbf{Turing}\cite{Turing} consists of Bing queries encoded by Turing AGI v5 that trains Transformers to capture similarity of intent in web search queries. We employed slices ranging in size from 1M to 100M for scalability evaluation.}
\item \textbf{Gist}\cite{laurent2010datasets} is an image dataset which contains about 1M data points with 960 dimensional features.
\item \textbf{Sift}\cite{laurent2010datasets} contains 1M SIFT vectors with 128 dimensions.
\item \textbf{Ukbench}\cite{UKBench} is a dataset that consists of about 1M images with 128 dimensional features.
\end{itemize}

For the training of our RPQ, similar to \cite{sablayrolles2018spreading, prokhorenkova2020graph, zhang2022connecting}, we configure the training set as a subset with 500K vectors and use the originally provided query set as the testing set.
%Due to the lack of training set, we segment the dataset and select portions that do not overlap with the base set as the training set.

\setlength{\textfloatsep}{0.15cm}
\begin{table}
\setlength{\abovecaptionskip}{0.1cm}
\centering
  \caption{\textbf{Dataset Statistics}}
  \label{tab:dataset}
  \renewcommand\arraystretch{1.3}
  \scalebox{0.9}{
  \begin{tabular}{c||c|c|c|c}
    \hline
    \textbf{Dataset}  & \textbf{Dimensions} & \textbf{Base} (M:$10^6$, B:$10^9$) & \textbf{Query}  & \textbf{LID}\cite{facco2017estimating}    \\
    \hline
    \hline
    BigANN \cite{baranchuk2021benchmarks}            & 128 & 1M$\sim$1B         & 10,000               & 16.6\\
    \hline   
    Deep \cite{YandexBenchmark}          & 96   & 1M$\sim$1B          & 10,000               & 17.6\\
    %\hline
    %Turing \cite{Turing}                              &100 & 1M$\sim$100M
            %& 10,000               & 30.5\\
    \hline
    Gist   \cite{laurent2010datasets}           & 960  & 1M       & 1,000                & 35.0\\
    \hline
    Sift \cite{laurent2010datasets}             & 128 & 1M        & 10,000                & 16.6                \\
    \hline
    Ukbench \cite{UKBench}        & 128     & 1M        & 200                &8.3\\
    \hline
    \end{tabular}
    }
    
\end{table}

\vspace{0.1cm}
\noindent \textbf{Comparing algorithms}. We first introduce three generic PQ methods: (1) \textbf{PQ} \cite{jegou2010product} is a typical quantization method that is used in DiskANN \cite{jayaram2019diskann} and other large indices \cite{baranchuk2018revisiting, johnson2019billion}. (2) \textbf{OPQ} \cite{ge2013optimized} optimizes PQ's vector division and is reported as a reliable quantization method \cite{echihabi2020return, blalock2017bolt}. (3) \textbf{Catalyst} \cite{sablayrolles2018spreading} utilizes a compression network to optimize quantization. Next, we integrate them and our RPQ with existing graph-based ANNS methods to evaluate their performance in hybrid (SSD+memory) and in-memory scenarios, as we discussed in \S \ref{sec:search}. It is worth mentioning that we provide memory constraints for both scenarios via Docker technology, more details are provided in the Resource constraints part below.

\begin{figure*}
\setlength{\abovecaptionskip}{0.1cm}
\centering
\includegraphics[width=179mm]{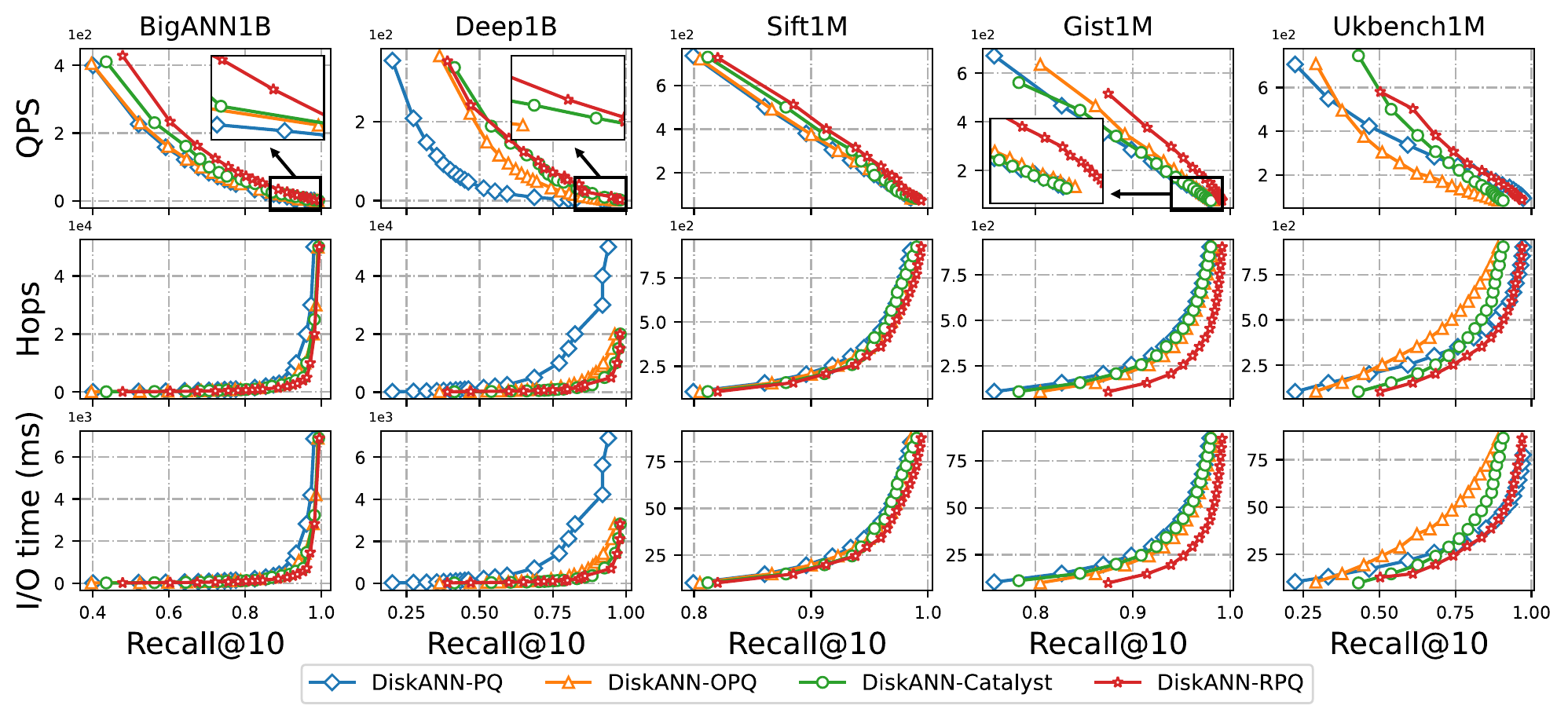}
\caption{\textbf{QPS, Hops, Disk I/O time vs. Recall@10 for different PQ-integrated graph-based ANNS (using PQ, OPQ, Catalyst and RPQ) in hybrid scenario (integrated with DiskANN). Each column reports the results for the same dataset.}}
\label{fig:QPS}
\vspace{-0.2cm}
\end{figure*}

\vspace{0.1cm}
\noindent\underline{Hybrid scenario (SSD+memory).} Since DiskANN and its variants are mainstream of this type of work, we establish comparing algorithms as follows. We retain the PG, i.e., Vamana, and original vectors in SSD, then we integrate four PQ methods with the same PG and vectors to form four algorithms. (1) \textbf{DiskANN-PQ} (the original DiskANN uses PQ \cite{jegou2010product} as default), (2) \textbf{DiskANN-OPQ}, (3) \textbf{DiskANN-Catalyst}, and (4) \textbf{DiskANN-RPQ}.

\vspace{0.1cm}
\noindent{\underline{In-memory scenario.} Since HNSW and NSG are reported as reliable graph-based ANNS algorithms in most of datasets \cite{Wang2021}, we evaluate the performance of PQ-integrated algorithms based on them. Note that, (5) \textbf{HNSW-PQ}\cite{faissweb, matsui2022arm, liu2023cmlocate} and (6) \textbf{L\&C}\cite{douze2018link, yang2021hierarchical} (L\&C uses a refined version of PQ \cite{jegou2010product}) are existing works that built atop HNSW, so we directly use them in our evaluation. Besides, we also establish (7) \textbf{HNSW-OPQ}, (8) \textbf{HNSW-Catalyst}, and (9) \textbf{HNSW-RPQ} for HNSW. Similar, we establish (10) \textbf{NSG-PQ}, (11) \textbf{NSG-OPQ}, (12) \textbf{NSG-Catalyst}, and (13) \textbf{NSG-RPQ} for NSG.

\vspace{0.1cm}
\noindent\textbf{Evaluation metrics.} We evaluate the search efficiency and accuracy with Queries Per Second (QPS) and Recall@$k$, which are widely used for graph-based ANNS \cite{fu2021high, douze2018link, Wang2021, jaiswal2022ood}. Specifically, QPS is the ratio of the number of queries to the search time, and Recall@$k$ is defined by Eq. \ref{eq:recall}. Besides, we introduce the number of routing hops, i.e., the number of next-hop selections during the routing process, and the average disk I/O time for a query as supplementary metrics to evaluate the search efficiency.

\vspace{0.1cm}
\noindent \textbf{Implementation setup}. The code of all comparing methods is publicly available in their respective GitHub repositories. All experiments were conducted on a Linux server with 8$\times$ NVIDIA Tesla V100, 2$\times$ Intel Xeon Processor (Skylake, IBRS) at 3.00GHz, and a 373G memory. 
% The PGs' build algorithms are implemented in C++ and compiled with g++ 7.5. The indexes are built with all 40 threads, but the search is evaluated on a single thread. All experiments are performed at least three times.

%Since the PG construction requires to load all the vectors in memory
\vspace{0.1cm}
\noindent \textbf{Resource constraints.} Similar to \cite{jayaram2019diskann,fu2017fast}, we use the full memory to build a one-shot PG for graph-based ANNS, i.e., Vamana for comparing methods (1)-(4) and HNSW, NSG for (5)-(13), and then we set the memory constraint via Docker to perform PQ-integrated graph-based ANNS. In DiskANN \cite{jayaram2019diskann}, it configures a fixed memory constraint as 64GB to record compact codes and codebook for all datasets. Such a configuration is unfair for some datasets. Suppose we set a fixed 64GB constraint, it is valid for BigANN-1B (600GB in total for PG and vectors), while it is invalid for BigANN-100M (60GB in total) as it can fit into 64GB memory. Different from \cite{jayaram2019diskann}, we set the memory constraint as a fixed fraction $f$ of the size of a dataset and graph. Here, we set $f=\frac{1}{32}\approx 3\%$ to get a more strict memory constraint than \cite{jayaram2019diskann}. For example, for 600GB BigANN-1B, we have the memory constraint as 18GB which is quite smaller than the fixed 64GB of \cite{jayaram2019diskann}. While for the 60GB BigANN-100M, we have the constraint as 1.8GB.

\vspace{0.1cm}
\noindent \textbf{Parameters}. For different PGs, we following the same procedure as \cite{Wang2021} to search for the optimal value of all the adjustable parameters, to make the algorithms' search performance reach the optimal level. The number of codewords in each sub-codebook is set to $K=$ 256, which enables the original vectors to be compactly encoded by several whole bytes. For the approximate process of Catalyst, we use the parameters as follows: $d_{out}=$ 40, $\lambda=$ 0.005 for 128 bits. For L\&C, we use the parameters as follows: $L=$ 8, $nsq=$ 1, $beta\_k=$ 1.

\vspace{-0.3cm}
\subsection{Efficiency and Effectiveness Evaluation} \label{sec:efficiency}

\noindent\textbf{SSD-memory hybrid scenario}. Figure \ref{fig:QPS} reports the efficiency (QPS, Hops, and Disk I/O time) vs. effectiveness (Recall@$10$) results for PQ-integrated graph-based ANNS using four state-of-the-art PQ methods, namely PQ, OPQ, and Catalyst atop DiskANN. Each column in Figure \ref{fig:QPS} corresponds to a different dataset, and we utilized the maximum scale for these datasets (1B for BigANN and Deep, 100M for Turing, and 1M for others). All searches were carried out using 8 threads, making full use of I/O resources. Our evaluations consistently demonstrate that DiskANN-RPQ outperforms competitors using other PQ methods with a better QPS vs. Recall@$10$ (the further to the upper right, the better the result). For instance, given the same Recall@$10$ at 95\% in Gist, DiskANN-RPQ achieves a QPS of 251.98, which is 77\% improvement (or 1.77$\times$ faster) w.r.t. that of DiskANN-PQ with a QPS of 142.3. The QPS improvement for BigANN, Deep, and SIFT are 135\%, 320\%, and 12\%, respectively. This can be explained that our RPQ considers both the neighborhood and routing features to learn a differential quantizer that is more fit to the graph-based ANNS. Besides, since RPQ adopts an adaptive vector decomposition to make imbalanced vector features uniformly distributed among all sub-vectors, DiskANN-RPQ can well-support the imbalanced datasets such as Gist and Deep.

\begin{figure*}
\setlength{\abovecaptionskip}{0.1cm}
\centering
\includegraphics[width=179mm]{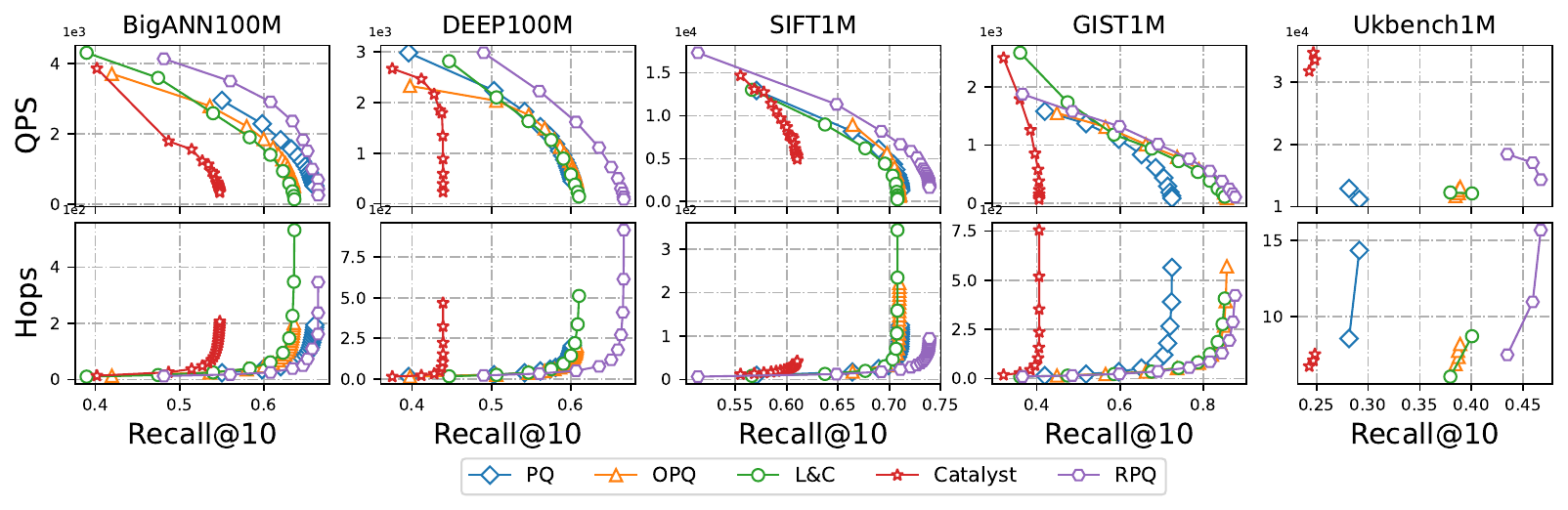}
%\vspace{-1.8cm}
\caption{\textbf{QPS and Hops vs. Recall@10 for in-memory scenario (HNSW as the default PG)}}\label{fig:6}
\vspace{-0.2cm}
\end{figure*}

\begin{figure*}
\setlength{\abovecaptionskip}{-0.2cm}
\centering
\includegraphics[width=179mm]{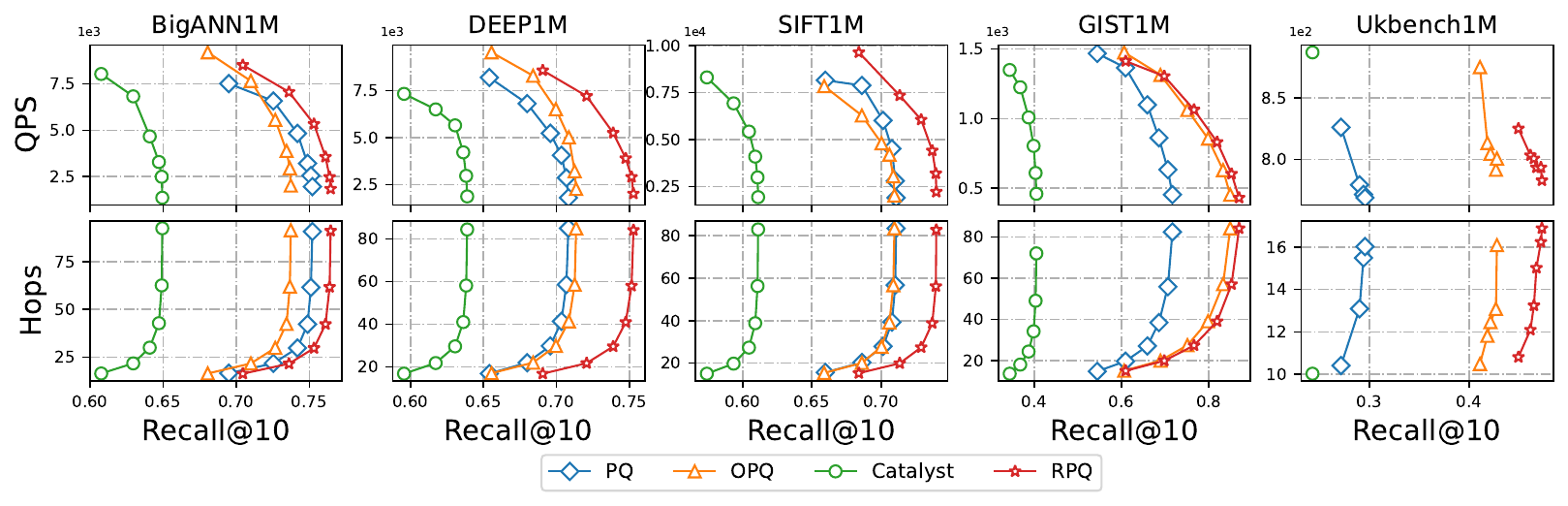}
\vspace{-0.1cm}
\caption{\textbf{QPS and Hops vs. Recall@10 for in-memory scenario (NSG as the default PG)}}\label{fig:7}
\vspace{-0.3cm}
\end{figure*}

Moreover, we provide the Hops vs. Recall@$10$ and Disk I/O time vs. Recall@$10$ over all datasets in Figure \ref{fig:QPS}. Note that, Hops increases as Recall@$10$ increases. This is because that we need more routing steps to retrieve more accurate results. The more the hops, the more the SSD accesses are required, resulting in an increasing Disk I/O time and decreasing QPS.

\vspace{0.1cm}
\noindent\textbf{In memory scenario.} As mentioned in \S \ref{sec:search}, the memory footprint of PQ-integrated graph-based ANNS for in-memory scenario mainly comes from the PG storage. And the PGs for two 1B datasets (BigANN 1B and Deep 1B) are too large to be maintained in memory with a strict resource constraint. So, we only provided the results for datasets up to 100M in Figure \ref{fig:6} and Figure \ref{fig:7}. For both HNSW and NSG, the integrated methods with our RPQ perform better than others, which are attributed to our consideration of two types features and losses.

\begin{table}[t]
\setlength{\abovecaptionskip}{0.1cm}
 \centering
  \caption{\textbf{Training time (hours)}}
  \label{tab:traincost}
  \renewcommand\arraystretch{1.3}
  \begin{tabular}{c||c|c|c|c|c}
    \hline
    \textbf{Methods} & \textbf{BigANN} & \textbf{Deep} & \textbf{Sift} & \textbf{Gist}  &  \textbf{Ukbench}  \\
    \hline
    \hline
    Catalyst            & 3.27 & 3.25 & 0.64 & 4.24 & 0.61\\
    \hline 
    RPQ            & 3.25 & 3.17 & 0.51 & 4.56 & 0.42\\
    \hline 
\end{tabular}
%\vspace{-0.2cm}
\end{table}

\begin{table}[t]
\setlength{\abovecaptionskip}{0.1cm}
  \centering
  \caption{\textbf{Model size (MB)}}
  \label{tab:modelsize}
  \renewcommand\arraystretch{1.3}
  \begin{tabular}{c||c|c|c|c|c}
    \hline
    \textbf{Methods} & \textbf{BigANN} & \textbf{Deep} &  \textbf{Sift} & \textbf{Gist}  &  \textbf{Ukbench}  \\
    \hline
    \hline
    Catalyst            & 4.7 & 4.6            & 4.7 & 8.3 & 4.7\\
    \hline 
    RPQ            & 0.69 & 0.46         & 0.78 & 1.8 & 0.69\\
    \hline 
\end{tabular}
\end{table}

\begin{table}[t]
\setlength{\abovecaptionskip}{0.1cm}
  \caption{\textbf{Effect of different features and losses used in RPQ for SSD-memory hybrid scenario}}
  \label{tab:ablation}
  \renewcommand\arraystretch{1.3}
  \begin{tabular}{c||c|c|c|c|c}
    \hline
    \textbf{Methods} & \textbf{BigANN} & \textbf{Deep} & \textbf{Gist} & \textbf{Sift} & \textbf{Ukbench}  \\
    \hline
    \hline
    RPQ            & 250.17 & 193.13   & 251.98  & 264.12 & 104.3   \\
    \hline 
    RPQ w/ N       & 231.00 & 174.02   & 228.42  & 250.41 & 101.92   \\
    \hline 
    RPQ w/ R       & 101.23 & 92.31    & 149.50  & 110.67 & 57.71   \\
    \hline 
    RPQ w/ L2R     & 80.21  & 77.41    & 70.14   & 92.37  & 56.36  \\
    \hline 
\end{tabular}
\end{table}

\begin{table}[t]
\setlength{\abovecaptionskip}{0.1cm}
  \caption{\textbf{Effect of different features and losses used in RPQ for in memory scenario}}
  \label{tab:ablation2}
  \renewcommand\arraystretch{1.3}
  \begin{tabular}{c||c|c|c|c|c}
    \hline
    \textbf{Methods} & \textbf{BigANN} & \textbf{Deep} & \textbf{Gist} & \textbf{Sift} & \textbf{Ukbench}  \\
    \hline
    \hline
    RPQ            & 5347.59 & 2906.97   & 7352.94  & 833.33 & 769.23   \\
    \hline 
    RPQ w/ N       & 5229.17 & 2517.59   & 6995.29  & 823.67 & 754.40   \\
    \hline 
    RPQ w/ R       & 3309.41 & 1827.36   & 4237.15  & 552.31 & 410.36   \\
    \hline 
    RPQ w/ L2R     & 3057.22 & 1554.09   & 3784.90  & 506.97 & 380.25  \\
    \hline 
\end{tabular}
\end{table}

\vspace{0.1cm}
\noindent\textbf{Training time and mode size.} We provided the training time (in hours) and the model size (in MB) of our RPQ and another leaning-based Catalyst over all datasets in Table \ref{tab:traincost} and Table \ref{tab:modelsize}. From the reported results, ours consumes a little more training time than Catalyst because we consider more features w.r.t. graph-based ANNS than Catalyst. Fortunately, we traded a small amount of additional overhead for better effectiveness and efficiency than Catalyst. Besides, for both two methods, the storage overhead for maintaining model is modest.

%\vspace{-0.1cm}
\subsection{Ablation Analysis} \label{sec:ablation}
\noindent\textbf{SSD-memory hybrid scenario.} We show the effects of different features and losses used in RPQ on ANNS's performance with following configurations: (1) RPQ with only neighborhood features and loss (RPQ w/ N), (2) RPQ with only routing features and loss (RPQ w/ R), (3) RPQ with two features and losses (RPQ), and (4) RPQ with L2R \cite{baranchuk2019learning} (RPQ w/ L2R). Table \ref{tab:ablation} shows the QPS results obtained at the same Recall@10 as 95\% for all datasets.  Note that, (3) performs the best and ours (1)-(3) are better than (4), this proves that our solution that considers both the neighborhood and routing features are necessary for achieving a good performance for PQ-integrated graph-based ANNS.

\begin{figure}[t]
\setlength{\abovecaptionskip}{0.1cm}
\centering
\includegraphics[width=85mm]{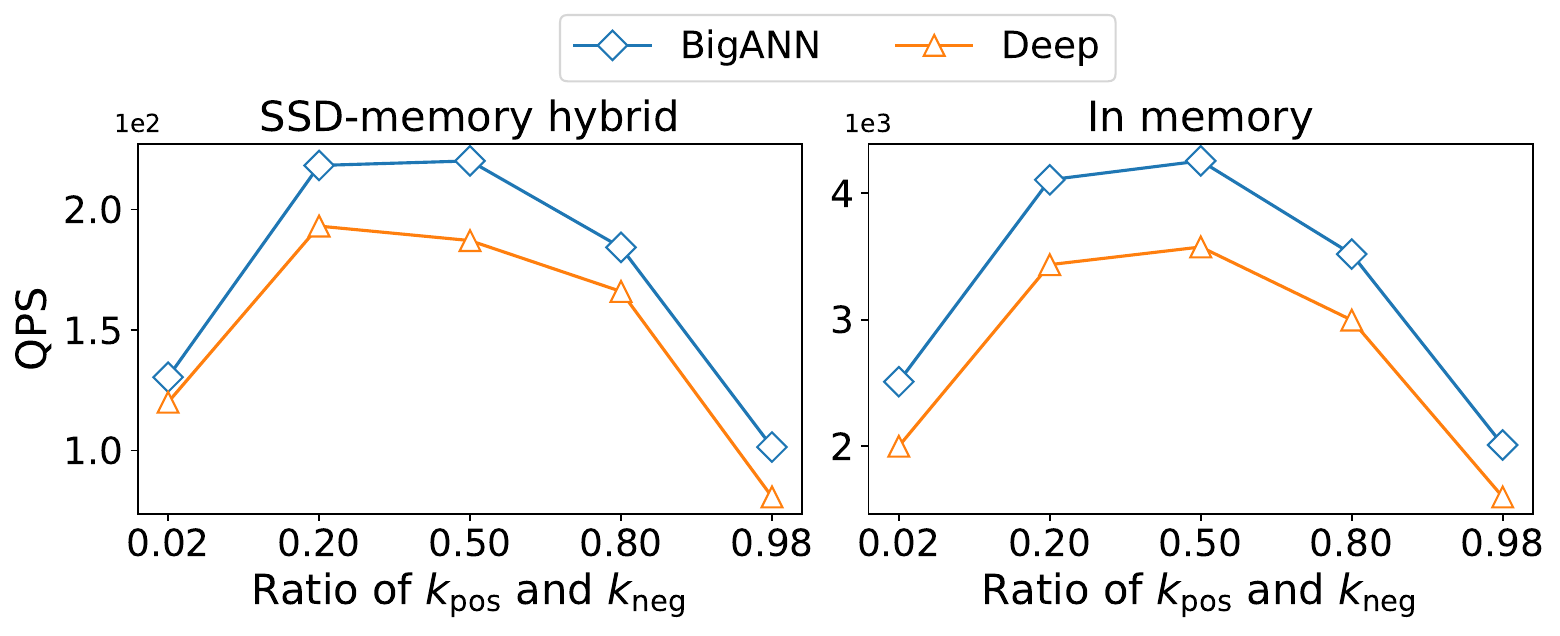}
\caption{Effect of $k_{\rm pos}$/$k_{\rm neg}$}\label{fig:sample}
% \vspace{-0.4cm}
\end{figure}

\noindent\textbf{In memory scenario.} As mentioned in Table \ref{tab:ablation2}, we also present the effects of different features and losses used in RPQ for in-memory scenario. Different with SSD-memory hybird scenario, we adopt the different evaluation standard for five datasets. For BigANN and DEEP, we use the QPS results obtained at the Recall@10 as 75\%. For SIFT, GIST and Ukbench, we use the QPS results obtained at the Recall@10 as 70\%, 80\% and 45\% respectively. Obviously, (3) also performs the best performance, this proves our solution can achieve a good performance for not only SSD-memory hybird scenario but also in memory scenario.

\vspace{-0.1cm}
\subsection{Parameter Sensitivity} \label{sec:parameter}
\noindent\underline{Effect of $k_{\rm pos}$ and $k_{\rm neg}$.} Since the proportion of positive and negative samples is critical for contrastive learning \cite{zhan2021optimizing}, we show the effect of $k_{\rm pos}/k_{\rm neg}$ on ANNS's performance over different datasets in Figure \ref{fig:sample} (using the QPS results achieved for the same Recall@10 as 95\%). A good QPS can be obtained when $k_{\rm pos}/k_{\rm neg}$ is configured in the range of [0.2,0.5].

\vspace{0.1cm}
\noindent\underline{Effect of $K$ and $M$.} We study the effect of $K$ and $M$ on ANNS's performance in Figure \ref{fig:k}, each value in a grid is a QPS obtained at the same Recall@10 as 95\% given a specific pair of $K$ and $M$. The larger the $K$ and $M$, the more the QPS we achieve. This can be explained that the larger the $K$, the more the codewords in a codebook, resulting in a more accurate PQ distance to a query. So, the routing process would converge quickly and lead to a larger QPS. Similarly, the larger the $M$, the more sub-vectors we have, so that the coding space is larger, making the PQ distance more accurate. For in memory scenario, we present the upper limit of Recall@10 in Figure \ref{fig:k2}, similar to SSD-memory hybrid scenario, the larget the K and M, the more the Recall@10 we can achieve.

\begin{figure}[t]
\setlength{\abovecaptionskip}{0.1cm}
\centering
\includegraphics[width=85mm]{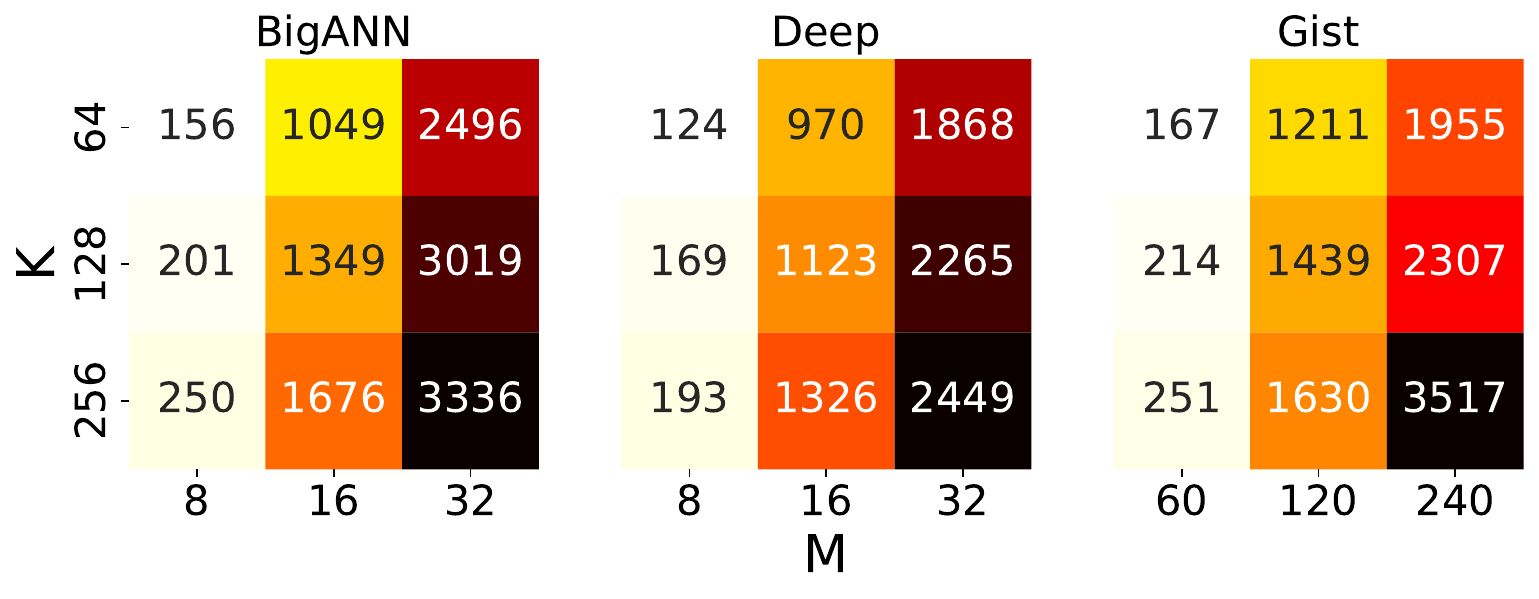}
\caption{Effect of $K$ and $M$ for SSD-memory hybrid scenario}\label{fig:k}
\end{figure}

\begin{figure}[t]
\setlength{\abovecaptionskip}{0.1cm}
\centering
\includegraphics[width=85mm]{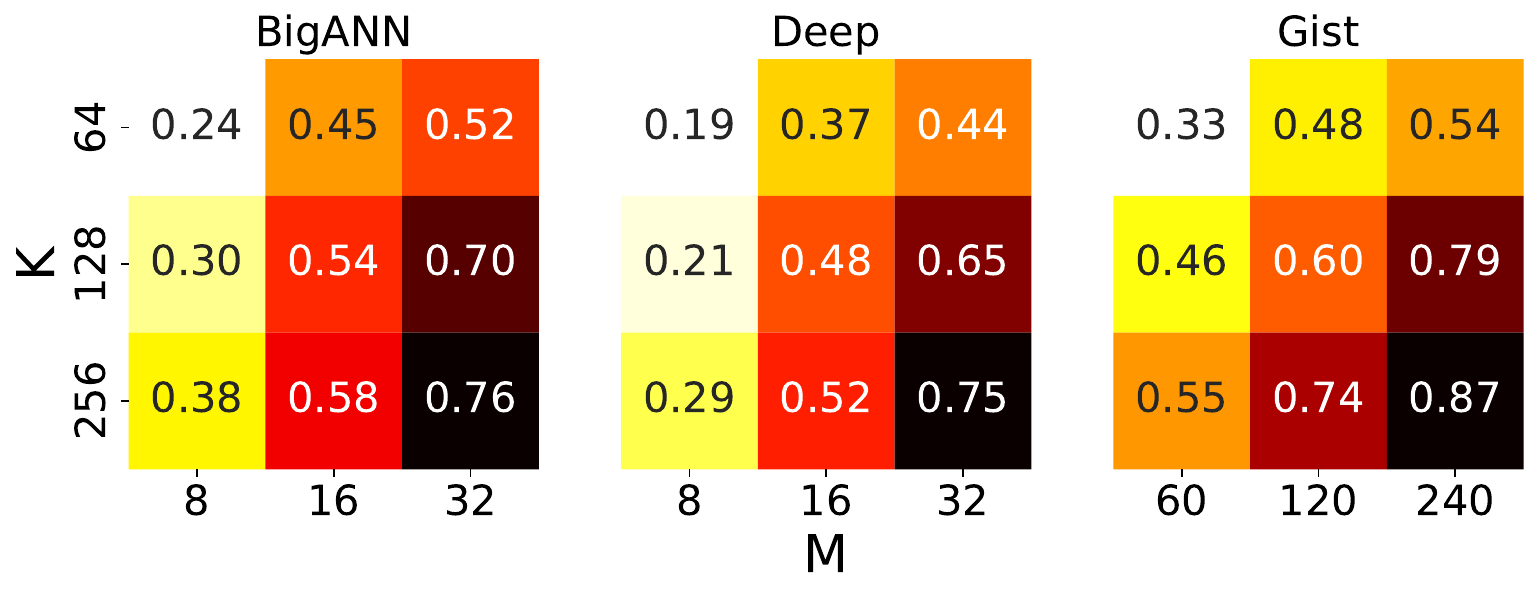}
\caption{Effect of $K$ and $M$ for in memory scenario}\label{fig:k2}
\end{figure}

\begin{figure}[h]
\centering
%\vspace{-0.4cm}
\includegraphics[width=80mm]{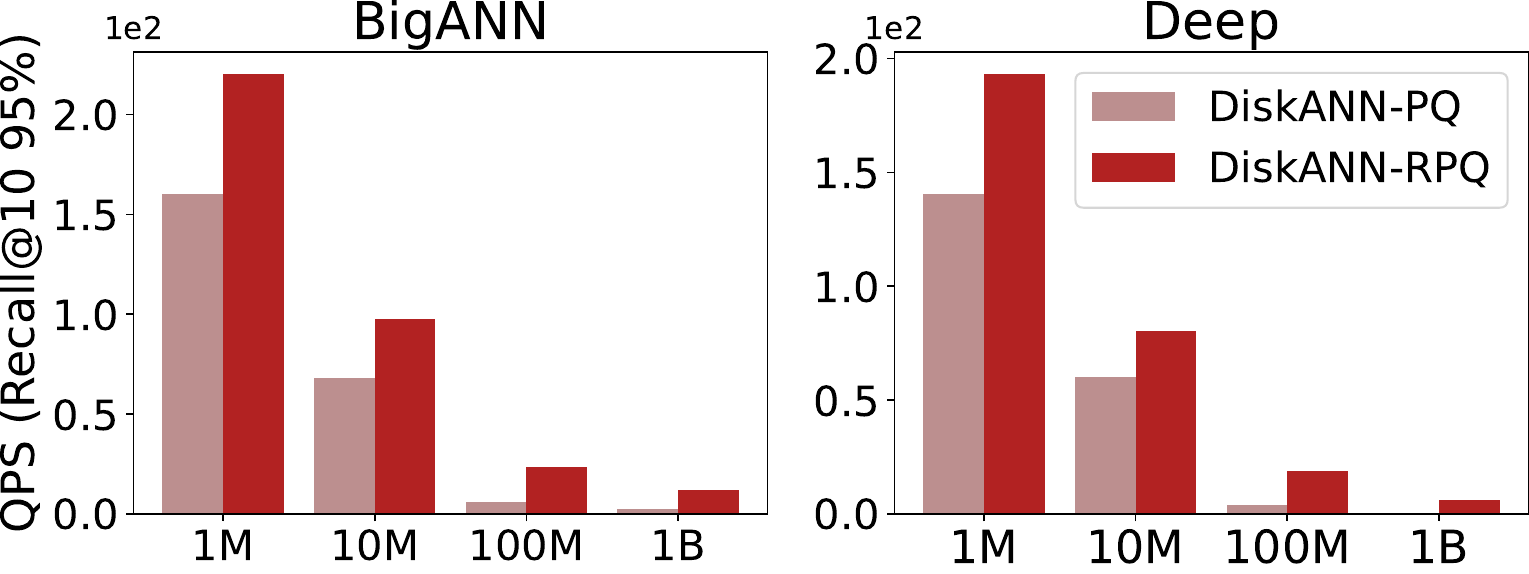}
\caption{Scalability analysis on data scales for SSD-memory hybrid scenario}\label{fig:scale}
%\vspace{-0.2cm}
\end{figure}

\begin{figure}[h]
\centering
%\vspace{-0.4cm}
\includegraphics[width=80mm]{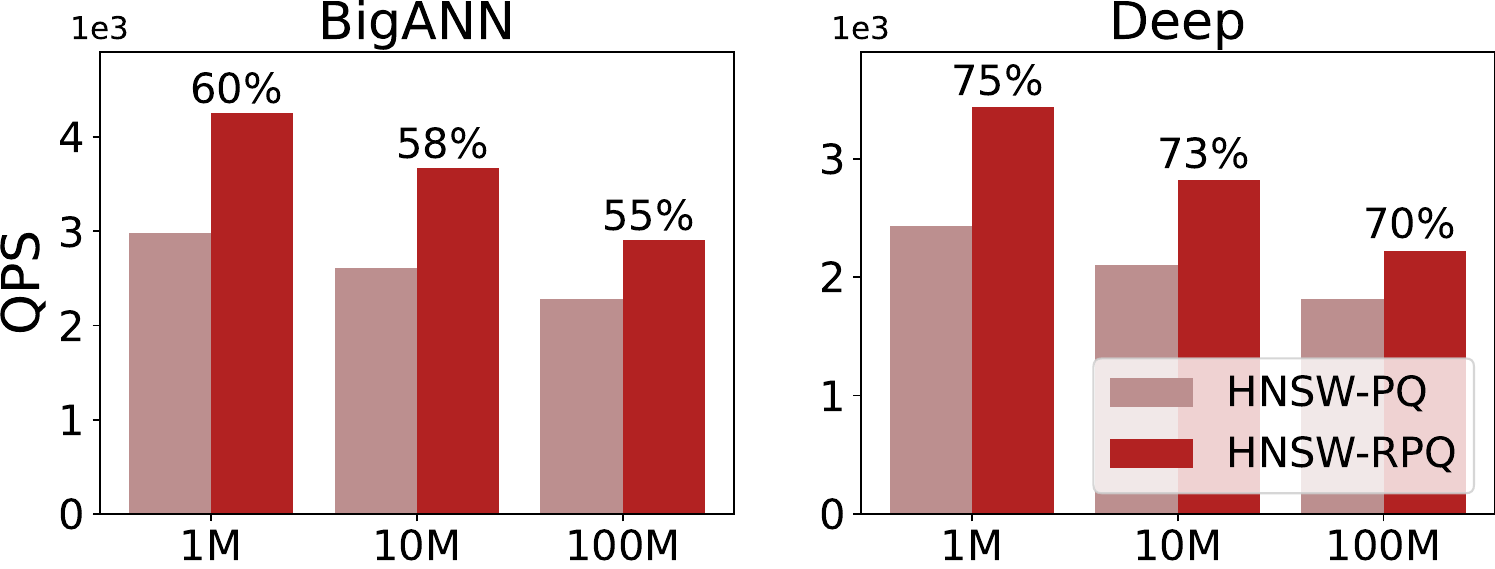}
\caption{Scalability analysis on data scales for in memory scenario}\label{fig:scale2}
%\vspace{-0.2cm}
\end{figure}

\vspace{-0.1cm}
\subsection{Scalability Analysis} \label{sec:scalability}
\noindent\textbf{SSD-memory hybrid scenario.} We show the scalability of various methods on different scales of BigANN and Deep datasets that varies from 1M to 1B. Each data point in Figure \ref{fig:scale} represents the QPS achieved at the same Recall@10 as 95\%. We found that our method outperforms others, showing a better scalability on scales.

\noindent\textbf{In memory scenario.} We present the scalability of HNSW-PQ and HNSW-RPQ on different scales of BigANN and Deep datasets that varies from 1M to 100M. Different with SSD-memory hybrid scenario, we represent the QPS achieved with various Recall@10 (denoted above the bar). We found that our method also outperforms others in memory scenario.

\section{Related Work} \label{sec:related}

We review some of the fundamental techniques related to our study. We first briefly overview several compression methods. Then, we list recent efforts in index-aware compression.

\subsection{Vector encoding methods}

% As very large datasets of high-dimensional vectors proliferate, ANNS rely on lossy compression or hashing schemes. A crucial requirement for these schemes is the ability to evaluate distances and scalar products between compressed and uncompressed vectors efficiently and without explicit decompression. 
Common  methods for high-dimentional compression mostly fall into two separate lines of research, binary and quantization methods. At the moment, systems that rely on the quantization compression are often preferred to hashing approaches due to a more favourable accuracy. Recent studies summing or concatenating codewords from several different codebooks and achieve advanced compression. Among them, PQ \cite{jegou2010product} is more simple and fast, which have been widespread adopted among high-tech companies nowadays. However, PQ implicitly relies on the limited amount of correlation between the dimension groups. To reduce the quantization distortions in PQ, some studies searched for the optimal decomposition to learn codebooks. OPQ \cite{ge2013optimized} improves PQ by finding better subspace partitioning and achieves a better search performance. However, these methods are not specifically designed for graphs and do not take into account the important routing processes in neighboring graphs. These routing information are considered important in multiple literature \cite{baranchuk2019learning, peng2022lan}.

\subsection{Index-aware compression}

The literature on index-aware compression is most relevant to our work, which can be traced back to this ground-breaking paper, which design and train a neural network that favor adapting quantizers to the index. Since then, index-aware compression approaches have attracted wide interests and shown exciting results, with plenty of ingeniously designed algorithms being developed. Among them, following the state-of-the-art PG, many studies try to adapt the encoders to the PG index. Same as the methods we talk above, URPH \cite{karaman2019unsupervised} propose an unsupervised hashing method, exploiting a shallow neural network, that aims to maintaining the effective search performance of HNSW by preserve the ranking induced by an original real-valued representation. CNT \cite{zhang2022connecting} proposed method consists of a compression network with transformer (CNT) which combines traditional projection function and transformer model, and an inhomogeneous neighborhood relationship preserving (INRP) loss which is aligned with the characteristic of ANNS. Link\&Code encode the index vectors using optimized product quantization and exploit the graph structure to refine the similarity estimation. It learns a regression codebook by alternate optimization to minimize the reconstruction error, which providing high precision with a small set of comparisons.

% Besides that, recently methods try to consider the graph structure into compression process. Link\&Code encode the index vectors using optimized product quantization and exploit the graph structure to refine the similarity estimation. It learns a regression codebook by alternate optimization to minimize the reconstruction error, which providing high precision with a relatively small set of comparisons. HVS adopt the hierarchical structure same as HNSW, which divided subspace divisions in a coarse-to-fine manner. 

% Learning2Index ranks clusters based on their nearest neighbor probabilities rather than the query-centroid distances. The nearest neighbor probabilities are estimated by employing neural networks to characterize the neighborhood relationships, i.e., the density function of nearest neighbors with respect to the query. The proposed probability-based ranking can replace the conventional distance-based ranking for finding candidate clusters, and the predicted probability can be used to determine the data quantity to be retrieved from the candidate cluster. 

% \begin{figure}
% \includegraphics[width=40mm]{pic/fig7.png}
% \caption{Overview of neural vector encoding pipeline of graph-based methods.}\label{fig:6}
% \end{figure}

\section{Conclusion} \label{sec:conclusion}

We studied the product quantization methods for graph-based approximate nearest neighbor search by considering the routing features. We first propose a general routing-guided PQ framework. Moreover, by using the RPQ framework, we next presented a differentiable quantizer which composed by adaptive vector decomposition and differentiable quantization. Besides, we extract the routing features as well as the neighborhood features using a sampling-based method. Finally, we take above neighborhood and routing features to train the differentiable quantizer, thus making the quantizer more adaptive to graph-based ANNS methods. Experimental results on real-world dataset confirmed the effectiveness and efficiency of our approach.

\section*{Acknowledgment}
This work was supported by the Primary R\&D Plan of Zhejiang (2021C03156 and 2023C03198) and the National NSF of China (62072149).

%\clearpage

\bibliographystyle{ACM-Reference-Format}
\bibliography{ref}


\begin{thebibliography}{75}

%%% the Web address.

\ifx \showCODEN    \undefined \def \showCODEN     #1{\unskip}     \fi
\ifx \showDOI      \undefined \def \showDOI       #1{#1}\fi
\ifx \showISBNx    \undefined \def \showISBNx     #1{\unskip}     \fi
\ifx \showISBNxiii \undefined \def \showISBNxiii  #1{\unskip}     \fi
\ifx \showISSN     \undefined \def \showISSN      #1{\unskip}     \fi
\ifx \showLCCN     \undefined \def \showLCCN      #1{\unskip}     \fi
\ifx \shownote     \undefined \def \shownote      #1{#1}          \fi
\ifx \showarticletitle \undefined \def \showarticletitle #1{#1}   \fi
\ifx \showURL      \undefined \def \showURL       {\relax}        \fi
% invisible to TeX
\providecommand\bibfield[2]{#2}
\providecommand\bibinfo[2]{#2}
\providecommand\natexlab[1]{#1}
\providecommand\showeprint[2][]{arXiv:#2}

\bibitem[\protect\citeauthoryear{??}{UKB}{Year}]%
        {UKBench}
 \bibinfo{year}{Year}\natexlab{}.
\newblock \bibinfo{title}{{UKBench Dataset}}.
\newblock \bibinfo{howpublished}{\url{https://archive.org/details/ukbench}}.
\newblock
\newblock
\shownote{Accessed on Date.}


\bibitem[\protect\citeauthoryear{Adeniyi, Wei, and Yongquan}{Adeniyi et~al\mbox{.}}{2016}]%
        {adeniyi2016automated}
\bibfield{author}{\bibinfo{person}{David~Adedayo Adeniyi}, \bibinfo{person}{Zhaoqiang Wei}, {and} \bibinfo{person}{Yang Yongquan}.} \bibinfo{year}{2016}\natexlab{}.
\newblock \showarticletitle{Automated web usage data mining and recommendation system using K-Nearest Neighbor (KNN) classification method}.
\newblock \bibinfo{journal}{\emph{Applied Computing and Informatics}} \bibinfo{volume}{12}, \bibinfo{number}{1} (\bibinfo{year}{2016}), \bibinfo{pages}{90--108}.
\newblock


\bibitem[\protect\citeauthoryear{Amsaleg, Chelly, Furon, Girard, Houle, Kawarabayashi, and Nett}{Amsaleg et~al\mbox{.}}{2015}]%
        {amsaleg2015estimating}
\bibfield{author}{\bibinfo{person}{Laurent Amsaleg}, \bibinfo{person}{Oussama Chelly}, \bibinfo{person}{Teddy Furon}, \bibinfo{person}{St{\'e}phane Girard}, \bibinfo{person}{Michael~E Houle}, \bibinfo{person}{Ken-ichi Kawarabayashi}, {and} \bibinfo{person}{Michael Nett}.} \bibinfo{year}{2015}\natexlab{}.
\newblock \showarticletitle{Estimating local intrinsic dimensionality}. In \bibinfo{booktitle}{\emph{Proceedings of the 21th ACM SIGKDD International Conference on Knowledge Discovery and Data Mining}}. \bibinfo{pages}{29--38}.
\newblock


\bibitem[\protect\citeauthoryear{Amsaleg and Jégou}{Amsaleg and Jégou}{2010}]%
        {laurent2010datasets}
\bibfield{author}{\bibinfo{person}{Laurent Amsaleg} {and} \bibinfo{person}{Hervé Jégou}.} \bibinfo{year}{2010}\natexlab{}.
\newblock \bibinfo{title}{Datasets for approximate nearest neighbor search}.
\newblock \bibinfo{howpublished}{Webpage}.
\newblock
\urldef\tempurl%
\url{http://corpus-texmex.irisa.fr}
\showURL{%
\tempurl}
\newblock
\shownote{Retrieved June 1, 2023.}


\bibitem[\protect\citeauthoryear{Amvrosiadis, Butt, Tarasov, Zadok, and Zhao}{Amvrosiadis et~al\mbox{.}}{2019}]%
        {amvrosiadis2019data}
\bibfield{author}{\bibinfo{person}{George Amvrosiadis}, \bibinfo{person}{Ali~R Butt}, \bibinfo{person}{Vasily Tarasov}, \bibinfo{person}{Erez Zadok}, {and} \bibinfo{person}{Ming Zhao}.} \bibinfo{year}{2019}\natexlab{}.
\newblock \showarticletitle{Data storage research vision 2025 report}.
\newblock \bibinfo{journal}{\emph{Technical Report}} (\bibinfo{year}{2019}).
\newblock


\bibitem[\protect\citeauthoryear{Aoyama, Ogawa, Hattori, Hori, and Nakamura}{Aoyama et~al\mbox{.}}{2013}]%
        {aoyama2013graph}
\bibfield{author}{\bibinfo{person}{Kazuo Aoyama}, \bibinfo{person}{Atsunori Ogawa}, \bibinfo{person}{Takashi Hattori}, \bibinfo{person}{Takaaki Hori}, {and} \bibinfo{person}{Atsushi Nakamura}.} \bibinfo{year}{2013}\natexlab{}.
\newblock \showarticletitle{Graph index based query-by-example search on a large speech data set}. In \bibinfo{booktitle}{\emph{2013 IEEE International Conference on Acoustics, Speech and Signal Processing}}. IEEE, \bibinfo{pages}{8520--8524}.
\newblock


\bibitem[\protect\citeauthoryear{Arora, Sinha, Kumar, and Bhattacharya}{Arora et~al\mbox{.}}{2018}]%
        {arora2018hd}
\bibfield{author}{\bibinfo{person}{Akhil Arora}, \bibinfo{person}{Sakshi Sinha}, \bibinfo{person}{Piyush Kumar}, {and} \bibinfo{person}{Arnab Bhattacharya}.} \bibinfo{year}{2018}\natexlab{}.
\newblock \showarticletitle{HD-Index: Pushing the Scalability-Accuracy Boundary for Approximate kNN Search in High-Dimensional Spaces}.
\newblock \bibinfo{journal}{\emph{Proceedings of the VLDB Endowment}} \bibinfo{volume}{11}, \bibinfo{number}{8} (\bibinfo{year}{2018}).
\newblock


\bibitem[\protect\citeauthoryear{Arya and Mount}{Arya and Mount}{1993}]%
        {arya1993approximate}
\bibfield{author}{\bibinfo{person}{Sunil Arya} {and} \bibinfo{person}{David~M Mount}.} \bibinfo{year}{1993}\natexlab{}.
\newblock \showarticletitle{Approximate nearest neighbor queries in fixed dimensions.}. In \bibinfo{booktitle}{\emph{SODA}}, Vol.~\bibinfo{volume}{93}. \bibinfo{pages}{271--280}.
\newblock


\bibitem[\protect\citeauthoryear{Arya, Mount, Netanyahu, Silverman, and Wu}{Arya et~al\mbox{.}}{1998}]%
        {arya1998optimal}
\bibfield{author}{\bibinfo{person}{Sunil Arya}, \bibinfo{person}{David~M Mount}, \bibinfo{person}{Nathan~S Netanyahu}, \bibinfo{person}{Ruth Silverman}, {and} \bibinfo{person}{Angela~Y Wu}.} \bibinfo{year}{1998}\natexlab{}.
\newblock \showarticletitle{An optimal algorithm for approximate nearest neighbor searching fixed dimensions}.
\newblock \bibinfo{journal}{\emph{Journal of the ACM (JACM)}} \bibinfo{volume}{45}, \bibinfo{number}{6} (\bibinfo{year}{1998}), \bibinfo{pages}{891--923}.
\newblock


\bibitem[\protect\citeauthoryear{Aum{\"u}ller, Bernhardsson, and Faithfull}{Aum{\"u}ller et~al\mbox{.}}{2020}]%
        {aumuller2020ann}
\bibfield{author}{\bibinfo{person}{Martin Aum{\"u}ller}, \bibinfo{person}{Erik Bernhardsson}, {and} \bibinfo{person}{Alexander Faithfull}.} \bibinfo{year}{2020}\natexlab{}.
\newblock \showarticletitle{ANN-Benchmarks: A benchmarking tool for approximate nearest neighbor algorithms}.
\newblock \bibinfo{journal}{\emph{Information Systems}}  \bibinfo{volume}{87} (\bibinfo{year}{2020}), \bibinfo{pages}{101374}.
\newblock


\bibitem[\protect\citeauthoryear{Baranchuk and Babenko}{Baranchuk and Babenko}{2021}]%
        {baranchuk2021benchmarks}
\bibfield{author}{\bibinfo{person}{Dmitry Baranchuk} {and} \bibinfo{person}{Artem Babenko}.} \bibinfo{year}{2021}\natexlab{}.
\newblock \bibinfo{title}{Benchmarks for Billion-Scale Similarity Search}.
\newblock \bibinfo{howpublished}{Webpage}.
\newblock
\urldef\tempurl%
\url{https://research.yandex.com/blog/benchmarks-for-billion-scale-similarity-search}
\showURL{%
\tempurl}
\newblock
\shownote{Retrieved June 1, 2023.}


\bibitem[\protect\citeauthoryear{Baranchuk, Babenko, and Malkov}{Baranchuk et~al\mbox{.}}{2018}]%
        {baranchuk2018revisiting}
\bibfield{author}{\bibinfo{person}{Dmitry Baranchuk}, \bibinfo{person}{Artem Babenko}, {and} \bibinfo{person}{Yury Malkov}.} \bibinfo{year}{2018}\natexlab{}.
\newblock \showarticletitle{Revisiting the inverted indices for billion-scale approximate nearest neighbors}. In \bibinfo{booktitle}{\emph{Proceedings of the European Conference on Computer Vision (ECCV)}}. \bibinfo{pages}{202--216}.
\newblock


\bibitem[\protect\citeauthoryear{Baranchuk, Persiyanov, Sinitsin, and Babenko}{Baranchuk et~al\mbox{.}}{2019}]%
        {baranchuk2019learning}
\bibfield{author}{\bibinfo{person}{Dmitry Baranchuk}, \bibinfo{person}{Dmitry Persiyanov}, \bibinfo{person}{Anton Sinitsin}, {and} \bibinfo{person}{Artem Babenko}.} \bibinfo{year}{2019}\natexlab{}.
\newblock \showarticletitle{Learning to Route in Similarity Graphs}. In \bibinfo{booktitle}{\emph{ICML}}. PMLR, \bibinfo{pages}{475--484}.
\newblock


\bibitem[\protect\citeauthoryear{Bijalwan, Kumar, Kumari, and Pascual}{Bijalwan et~al\mbox{.}}{2014}]%
        {bijalwan2014knn}
\bibfield{author}{\bibinfo{person}{Vishwanath Bijalwan}, \bibinfo{person}{Vinay Kumar}, \bibinfo{person}{Pinki Kumari}, {and} \bibinfo{person}{Jordan Pascual}.} \bibinfo{year}{2014}\natexlab{}.
\newblock \showarticletitle{KNN based machine learning approach for text and document mining}.
\newblock \bibinfo{journal}{\emph{International Journal of Database Theory and Application}} \bibinfo{volume}{7}, \bibinfo{number}{1} (\bibinfo{year}{2014}), \bibinfo{pages}{61--70}.
\newblock


\bibitem[\protect\citeauthoryear{Blalock and Guttag}{Blalock and Guttag}{2017}]%
        {blalock2017bolt}
\bibfield{author}{\bibinfo{person}{Davis~W Blalock} {and} \bibinfo{person}{John~V Guttag}.} \bibinfo{year}{2017}\natexlab{}.
\newblock \showarticletitle{Bolt: Accelerated data mining with fast vector compression}. In \bibinfo{booktitle}{\emph{Proceedings of the 23rd ACM SIGKDD International Conference on Knowledge Discovery and Data Mining}}. \bibinfo{pages}{727--735}.
\newblock


\bibitem[\protect\citeauthoryear{Chen, Zhao, Wang, Li, Liu, Li, Yang, and Wang}{Chen et~al\mbox{.}}{2021}]%
        {chen2021spann}
\bibfield{author}{\bibinfo{person}{Qi Chen}, \bibinfo{person}{Bing Zhao}, \bibinfo{person}{Haidong Wang}, \bibinfo{person}{Mingqin Li}, \bibinfo{person}{Chuanjie Liu}, \bibinfo{person}{Zengzhong Li}, \bibinfo{person}{Mao Yang}, {and} \bibinfo{person}{Jingdong Wang}.} \bibinfo{year}{2021}\natexlab{}.
\newblock \showarticletitle{SPANN: Highly-efficient Billion-scale Approximate Nearest Neighborhood Search}.
\newblock \bibinfo{journal}{\emph{Advances in Neural Information Processing Systems}}  \bibinfo{volume}{34} (\bibinfo{year}{2021}), \bibinfo{pages}{5199--5212}.
\newblock


\bibitem[\protect\citeauthoryear{Cohan, Feldman, Beltagy, Downey, and Weld}{Cohan et~al\mbox{.}}{2020}]%
        {Cohan2020}
\bibfield{author}{\bibinfo{person}{Arman Cohan}, \bibinfo{person}{Sergey Feldman}, \bibinfo{person}{Iz Beltagy}, \bibinfo{person}{Doug Downey}, {and} \bibinfo{person}{Daniel~S. Weld}.} \bibinfo{year}{2020}\natexlab{}.
\newblock \showarticletitle{{SPECTER:} Document-level Representation Learning using Citation-informed Transformers}. In \bibinfo{booktitle}{\emph{Proceedings of the 58th Annual Meeting of the Association for Computational Linguistics, {ACL} 2020, Online, July 5-10, 2020}}, \bibfield{editor}{\bibinfo{person}{Dan Jurafsky}, \bibinfo{person}{Joyce Chai}, \bibinfo{person}{Natalie Schluter}, {and} \bibinfo{person}{Joel~R. Tetreault}} (Eds.). \bibinfo{publisher}{Association for Computational Linguistics}, \bibinfo{pages}{2270--2282}.
\newblock
\urldef\tempurl%
\url{https://doi.org/10.18653/v1/2020.acl-main.207}
\showDOI{\tempurl}


\bibitem[\protect\citeauthoryear{Cover and Hart}{Cover and Hart}{1967}]%
        {cover1967nearest}
\bibfield{author}{\bibinfo{person}{Thomas Cover} {and} \bibinfo{person}{Peter Hart}.} \bibinfo{year}{1967}\natexlab{}.
\newblock \showarticletitle{Nearest neighbor pattern classification}.
\newblock \bibinfo{journal}{\emph{IEEE transactions on information theory}} \bibinfo{volume}{13}, \bibinfo{number}{1} (\bibinfo{year}{1967}), \bibinfo{pages}{21--27}.
\newblock


\bibitem[\protect\citeauthoryear{Datta, Joshi, Li, and Wang}{Datta et~al\mbox{.}}{2008}]%
        {datta2008image}
\bibfield{author}{\bibinfo{person}{Ritendra Datta}, \bibinfo{person}{Dhiraj Joshi}, \bibinfo{person}{Jia Li}, {and} \bibinfo{person}{James~Z Wang}.} \bibinfo{year}{2008}\natexlab{}.
\newblock \showarticletitle{Image retrieval: Ideas, influences, and trends of the new age}.
\newblock \bibinfo{journal}{\emph{ACM Computing Surveys (Csur)}} \bibinfo{volume}{40}, \bibinfo{number}{2} (\bibinfo{year}{2008}), \bibinfo{pages}{1--60}.
\newblock


\bibitem[\protect\citeauthoryear{Dobson, Shen, Blelloch, Dhulipala, Gu, Simhadri, and Sun}{Dobson et~al\mbox{.}}{2023}]%
        {dobson2023scaling}
\bibfield{author}{\bibinfo{person}{Magdalen Dobson}, \bibinfo{person}{Zheqi Shen}, \bibinfo{person}{Guy~E Blelloch}, \bibinfo{person}{Laxman Dhulipala}, \bibinfo{person}{Yan Gu}, \bibinfo{person}{Harsha~Vardhan Simhadri}, {and} \bibinfo{person}{Yihan Sun}.} \bibinfo{year}{2023}\natexlab{}.
\newblock \showarticletitle{Scaling Graph-Based ANNS Algorithms to Billion-Size Datasets: A Comparative Analysis}.
\newblock \bibinfo{journal}{\emph{arXiv preprint arXiv:2305.04359}} (\bibinfo{year}{2023}).
\newblock


\bibitem[\protect\citeauthoryear{Douze, Sablayrolles, and J{\'e}gou}{Douze et~al\mbox{.}}{2018}]%
        {douze2018link}
\bibfield{author}{\bibinfo{person}{Matthijs Douze}, \bibinfo{person}{Alexandre Sablayrolles}, {and} \bibinfo{person}{Herv{\'e} J{\'e}gou}.} \bibinfo{year}{2018}\natexlab{}.
\newblock \showarticletitle{Link and code: Fast indexing with graphs and compact regression codes}. In \bibinfo{booktitle}{\emph{Proceedings of the IEEE conference on computer vision and pattern recognition}}. \bibinfo{pages}{3646--3654}.
\newblock


\bibitem[\protect\citeauthoryear{Echihabi, Zoumpatianos, Palpanas, and Benbrahim}{Echihabi et~al\mbox{.}}{2020}]%
        {echihabi2020return}
\bibfield{author}{\bibinfo{person}{Karima Echihabi}, \bibinfo{person}{Kostas Zoumpatianos}, \bibinfo{person}{Themis Palpanas}, {and} \bibinfo{person}{Houda Benbrahim}.} \bibinfo{year}{2020}\natexlab{}.
\newblock \showarticletitle{Return of the lernaean hydra: Experimental evaluation of data series approximate similarity search}.
\newblock \bibinfo{journal}{\emph{arXiv preprint arXiv:2006.11459}} (\bibinfo{year}{2020}).
\newblock


\bibitem[\protect\citeauthoryear{Facco, d’Errico, Rodriguez, and Laio}{Facco et~al\mbox{.}}{2017}]%
        {facco2017estimating}
\bibfield{author}{\bibinfo{person}{Elena Facco}, \bibinfo{person}{Maria d’Errico}, \bibinfo{person}{Alex Rodriguez}, {and} \bibinfo{person}{Alessandro Laio}.} \bibinfo{year}{2017}\natexlab{}.
\newblock \showarticletitle{Estimating the intrinsic dimension of datasets by a minimal neighborhood information}.
\newblock \bibinfo{journal}{\emph{Scientific reports}} \bibinfo{volume}{7}, \bibinfo{number}{1} (\bibinfo{year}{2017}), \bibinfo{pages}{12140}.
\newblock


\bibitem[\protect\citeauthoryear{Flickner, Sawhney, Niblack, Ashley, Huang, Dom, Gorkani, Hafner, Lee, Petkovic, et~al\mbox{.}}{Flickner et~al\mbox{.}}{1995}]%
        {flickner1995query}
\bibfield{author}{\bibinfo{person}{Myron Flickner}, \bibinfo{person}{Harpreet Sawhney}, \bibinfo{person}{Wayne Niblack}, \bibinfo{person}{Jonathan Ashley}, \bibinfo{person}{Qian Huang}, \bibinfo{person}{Byron Dom}, \bibinfo{person}{Monika Gorkani}, \bibinfo{person}{Jim Hafner}, \bibinfo{person}{Denis Lee}, \bibinfo{person}{Dragutin Petkovic}, {et~al\mbox{.}}} \bibinfo{year}{1995}\natexlab{}.
\newblock \showarticletitle{Query by image and video content: The QBIC system}.
\newblock \bibinfo{journal}{\emph{computer}} \bibinfo{volume}{28}, \bibinfo{number}{9} (\bibinfo{year}{1995}), \bibinfo{pages}{23--32}.
\newblock


\bibitem[\protect\citeauthoryear{Fu, Wang, and Cai}{Fu et~al\mbox{.}}{2021}]%
        {fu2021high}
\bibfield{author}{\bibinfo{person}{Cong Fu}, \bibinfo{person}{Changxu Wang}, {and} \bibinfo{person}{Deng Cai}.} \bibinfo{year}{2021}\natexlab{}.
\newblock \showarticletitle{High dimensional similarity search with satellite system graph: Efficiency, scalability, and unindexed query compatibility}.
\newblock \bibinfo{journal}{\emph{IEEE Transactions on Pattern Analysis and Machine Intelligence}} \bibinfo{volume}{44}, \bibinfo{number}{8} (\bibinfo{year}{2021}), \bibinfo{pages}{4139--4150}.
\newblock


\bibitem[\protect\citeauthoryear{Fu, Xiang, Wang, and Cai}{Fu et~al\mbox{.}}{2017}]%
        {fu2017fast}
\bibfield{author}{\bibinfo{person}{Cong Fu}, \bibinfo{person}{Chao Xiang}, \bibinfo{person}{Changxu Wang}, {and} \bibinfo{person}{Deng Cai}.} \bibinfo{year}{2017}\natexlab{}.
\newblock \showarticletitle{Fast approximate nearest neighbor search with the navigating spreading-out graph}.
\newblock \bibinfo{journal}{\emph{arXiv preprint arXiv:1707.00143}} (\bibinfo{year}{2017}).
\newblock


\bibitem[\protect\citeauthoryear{Ge, He, Ke, and Sun}{Ge et~al\mbox{.}}{2013}]%
        {ge2013optimized}
\bibfield{author}{\bibinfo{person}{Tiezheng Ge}, \bibinfo{person}{Kaiming He}, \bibinfo{person}{Qifa Ke}, {and} \bibinfo{person}{Jian Sun}.} \bibinfo{year}{2013}\natexlab{}.
\newblock \showarticletitle{Optimized product quantization for approximate nearest neighbor search}. In \bibinfo{booktitle}{\emph{Proceedings of the IEEE Conference on Computer Vision and Pattern Recognition}}. \bibinfo{pages}{2946--2953}.
\newblock


\bibitem[\protect\citeauthoryear{Gollapudi, Karia, Sivashankar, Krishnaswamy, Begwani, Raz, Lin, Zhang, Mahapatro, Srinivasan, et~al\mbox{.}}{Gollapudi et~al\mbox{.}}{2023}]%
        {gollapudi2023filtered}
\bibfield{author}{\bibinfo{person}{Siddharth Gollapudi}, \bibinfo{person}{Neel Karia}, \bibinfo{person}{Varun Sivashankar}, \bibinfo{person}{Ravishankar Krishnaswamy}, \bibinfo{person}{Nikit Begwani}, \bibinfo{person}{Swapnil Raz}, \bibinfo{person}{Yiyong Lin}, \bibinfo{person}{Yin Zhang}, \bibinfo{person}{Neelam Mahapatro}, \bibinfo{person}{Premkumar Srinivasan}, {et~al\mbox{.}}} \bibinfo{year}{2023}\natexlab{}.
\newblock \showarticletitle{Filtered-DiskANN: Graph Algorithms for Approximate Nearest Neighbor Search with Filters}. In \bibinfo{booktitle}{\emph{Proceedings of the ACM Web Conference 2023}}. \bibinfo{pages}{3406--3416}.
\newblock


\bibitem[\protect\citeauthoryear{Guo, Luan, Xiang, Yan, Yi, Luo, Cheng, Xu, Luo, Liu, et~al\mbox{.}}{Guo et~al\mbox{.}}{2022}]%
        {guo2022manu}
\bibfield{author}{\bibinfo{person}{Rentong Guo}, \bibinfo{person}{Xiaofan Luan}, \bibinfo{person}{Long Xiang}, \bibinfo{person}{Xiao Yan}, \bibinfo{person}{Xiaomeng Yi}, \bibinfo{person}{Jigao Luo}, \bibinfo{person}{Qianya Cheng}, \bibinfo{person}{Weizhi Xu}, \bibinfo{person}{Jiarui Luo}, \bibinfo{person}{Frank Liu}, {et~al\mbox{.}}} \bibinfo{year}{2022}\natexlab{}.
\newblock \showarticletitle{Manu: a cloud native vector database management system}.
\newblock \bibinfo{journal}{\emph{arXiv preprint arXiv:2206.13843}} (\bibinfo{year}{2022}).
\newblock


\bibitem[\protect\citeauthoryear{Hacid and Yoshida}{Hacid and Yoshida}{2010}]%
        {hacid2010neighborhood}
\bibfield{author}{\bibinfo{person}{Hakim Hacid} {and} \bibinfo{person}{Tetsuya Yoshida}.} \bibinfo{year}{2010}\natexlab{}.
\newblock \showarticletitle{Neighborhood graphs for indexing and retrieving multi-dimensional data}.
\newblock \bibinfo{journal}{\emph{Journal of Intelligent Information Systems}}  \bibinfo{volume}{34} (\bibinfo{year}{2010}), \bibinfo{pages}{93--111}.
\newblock


\bibitem[\protect\citeauthoryear{He, Kumar, and Chang}{He et~al\mbox{.}}{2012}]%
        {he2012difficulty}
\bibfield{author}{\bibinfo{person}{Junfeng He}, \bibinfo{person}{Sanjiv Kumar}, {and} \bibinfo{person}{Shih-Fu Chang}.} \bibinfo{year}{2012}\natexlab{}.
\newblock \showarticletitle{On the difficulty of nearest neighbor search}.
\newblock \bibinfo{journal}{\emph{arXiv preprint arXiv:1206.6411}} (\bibinfo{year}{2012}).
\newblock


\bibitem[\protect\citeauthoryear{Huang, Feng, Fang, Ng, and Wang}{Huang et~al\mbox{.}}{2017}]%
        {huang2017query}
\bibfield{author}{\bibinfo{person}{Qiang Huang}, \bibinfo{person}{Jianlin Feng}, \bibinfo{person}{Qiong Fang}, \bibinfo{person}{Wilfred Ng}, {and} \bibinfo{person}{Wei Wang}.} \bibinfo{year}{2017}\natexlab{}.
\newblock \showarticletitle{Query-aware locality-sensitive hashing scheme for lp norm}.
\newblock \bibinfo{journal}{\emph{The VLDB Journal}} \bibinfo{volume}{26}, \bibinfo{number}{5} (\bibinfo{year}{2017}), \bibinfo{pages}{683--708}.
\newblock


\bibitem[\protect\citeauthoryear{Indyk and Motwani}{Indyk and Motwani}{1998}]%
        {indyk1998approximate}
\bibfield{author}{\bibinfo{person}{Piotr Indyk} {and} \bibinfo{person}{Rajeev Motwani}.} \bibinfo{year}{1998}\natexlab{}.
\newblock \showarticletitle{Approximate nearest neighbors: towards removing theƒƒ curse of dimensionality}. In \bibinfo{booktitle}{\emph{Proceedings of the thirtieth annual ACM symposium on Theory of computing}}. \bibinfo{pages}{604--613}.
\newblock


\bibitem[\protect\citeauthoryear{Jaiswal, Krishnaswamy, Garg, Simhadri, and Agrawal}{Jaiswal et~al\mbox{.}}{2022}]%
        {jaiswal2022ood}
\bibfield{author}{\bibinfo{person}{Shikhar Jaiswal}, \bibinfo{person}{Ravishankar Krishnaswamy}, \bibinfo{person}{Ankit Garg}, \bibinfo{person}{Harsha~Vardhan Simhadri}, {and} \bibinfo{person}{Sheshansh Agrawal}.} \bibinfo{year}{2022}\natexlab{}.
\newblock \showarticletitle{OOD-DiskANN: Efficient and Scalable Graph ANNS for Out-of-Distribution Queries}.
\newblock \bibinfo{journal}{\emph{arXiv preprint arXiv:2211.12850}} (\bibinfo{year}{2022}).
\newblock


\bibitem[\protect\citeauthoryear{Jang, Gu, and Poole}{Jang et~al\mbox{.}}{2016}]%
        {jang2016categorical}
\bibfield{author}{\bibinfo{person}{Eric Jang}, \bibinfo{person}{Shixiang Gu}, {and} \bibinfo{person}{Ben Poole}.} \bibinfo{year}{2016}\natexlab{}.
\newblock \showarticletitle{Categorical reparameterization with gumbel-softmax}.
\newblock \bibinfo{journal}{\emph{arXiv preprint arXiv:1611.01144}} (\bibinfo{year}{2016}).
\newblock


\bibitem[\protect\citeauthoryear{Jayaram~Subramanya, Devvrit, Simhadri, Krishnawamy, and Kadekodi}{Jayaram~Subramanya et~al\mbox{.}}{2019}]%
        {jayaram2019diskann}
\bibfield{author}{\bibinfo{person}{Suhas Jayaram~Subramanya}, \bibinfo{person}{Fnu Devvrit}, \bibinfo{person}{Harsha~Vardhan Simhadri}, \bibinfo{person}{Ravishankar Krishnawamy}, {and} \bibinfo{person}{Rohan Kadekodi}.} \bibinfo{year}{2019}\natexlab{}.
\newblock \showarticletitle{Diskann: Fast accurate billion-point nearest neighbor search on a single node}.
\newblock \bibinfo{journal}{\emph{Advances in Neural Information Processing Systems}}  \bibinfo{volume}{32} (\bibinfo{year}{2019}).
\newblock


\bibitem[\protect\citeauthoryear{Jegou, Douze, and Schmid}{Jegou et~al\mbox{.}}{2010}]%
        {jegou2010product}
\bibfield{author}{\bibinfo{person}{Herve Jegou}, \bibinfo{person}{Matthijs Douze}, {and} \bibinfo{person}{Cordelia Schmid}.} \bibinfo{year}{2010}\natexlab{}.
\newblock \showarticletitle{Product quantization for nearest neighbor search}.
\newblock \bibinfo{journal}{\emph{IEEE transactions on pattern analysis and machine intelligence}} \bibinfo{volume}{33}, \bibinfo{number}{1} (\bibinfo{year}{2010}), \bibinfo{pages}{117--128}.
\newblock


\bibitem[\protect\citeauthoryear{Johnson, Douze, and J{\'e}gou}{Johnson et~al\mbox{.}}{2019}]%
        {johnson2019billion}
\bibfield{author}{\bibinfo{person}{Jeff Johnson}, \bibinfo{person}{Matthijs Douze}, {and} \bibinfo{person}{Herv{\'e} J{\'e}gou}.} \bibinfo{year}{2019}\natexlab{}.
\newblock \showarticletitle{Billion-scale similarity search with {GPUs}}.
\newblock \bibinfo{journal}{\emph{IEEE Transactions on Big Data}} \bibinfo{volume}{7}, \bibinfo{number}{3} (\bibinfo{year}{2019}), \bibinfo{pages}{535--547}.
\newblock


\bibitem[\protect\citeauthoryear{Karaman, Lin, Hu, and Chang}{Karaman et~al\mbox{.}}{2019}]%
        {karaman2019unsupervised}
\bibfield{author}{\bibinfo{person}{Svebor Karaman}, \bibinfo{person}{Xudong Lin}, \bibinfo{person}{Xuefeng Hu}, {and} \bibinfo{person}{Shih-Fu Chang}.} \bibinfo{year}{2019}\natexlab{}.
\newblock \showarticletitle{Unsupervised rank-preserving hashing for large-scale image retrieval}. In \bibinfo{booktitle}{\emph{Proceedings of the 2019 on International Conference on Multimedia Retrieval}}. \bibinfo{pages}{192--196}.
\newblock


\bibitem[\protect\citeauthoryear{Kingma and Ba}{Kingma and Ba}{2014}]%
        {kingma2014adam}
\bibfield{author}{\bibinfo{person}{Diederik~P Kingma} {and} \bibinfo{person}{Jimmy Ba}.} \bibinfo{year}{2014}\natexlab{}.
\newblock \showarticletitle{Adam: A method for stochastic optimization}.
\newblock \bibinfo{journal}{\emph{arXiv preprint arXiv:1412.6980}} (\bibinfo{year}{2014}).
\newblock


\bibitem[\protect\citeauthoryear{Kosuge and Oshima}{Kosuge and Oshima}{2019}]%
        {kosuge2019object}
\bibfield{author}{\bibinfo{person}{Atsutake Kosuge} {and} \bibinfo{person}{Takashi Oshima}.} \bibinfo{year}{2019}\natexlab{}.
\newblock \showarticletitle{An object-pose estimation acceleration technique for picking robot applications by using graph-reusing k-nn search}. In \bibinfo{booktitle}{\emph{2019 First International Conference on Graph Computing (GC)}}. IEEE, \bibinfo{pages}{68--74}.
\newblock


\bibitem[\protect\citeauthoryear{Li, Zhang, Sun, Wang, Li, Zhang, and Lin}{Li et~al\mbox{.}}{2019}]%
        {li2019approximate}
\bibfield{author}{\bibinfo{person}{Wen Li}, \bibinfo{person}{Ying Zhang}, \bibinfo{person}{Yifang Sun}, \bibinfo{person}{Wei Wang}, \bibinfo{person}{Mingjie Li}, \bibinfo{person}{Wenjie Zhang}, {and} \bibinfo{person}{Xuemin Lin}.} \bibinfo{year}{2019}\natexlab{}.
\newblock \showarticletitle{Approximate nearest neighbor search on high dimensional data—experiments, analyses, and improvement}.
\newblock \bibinfo{journal}{\emph{IEEE Transactions on Knowledge and Data Engineering}} \bibinfo{volume}{32}, \bibinfo{number}{8} (\bibinfo{year}{2019}), \bibinfo{pages}{1475--1488}.
\newblock


\bibitem[\protect\citeauthoryear{Liu, Zhu, Hu, Sun, Liu, Liu, Dai, Yang, and Wang}{Liu et~al\mbox{.}}{2022}]%
        {liu2022optimizing}
\bibfield{author}{\bibinfo{person}{Jun Liu}, \bibinfo{person}{Zhenhua Zhu}, \bibinfo{person}{Jingbo Hu}, \bibinfo{person}{Hanbo Sun}, \bibinfo{person}{Li Liu}, \bibinfo{person}{Lingzhi Liu}, \bibinfo{person}{Guohao Dai}, \bibinfo{person}{Huazhong Yang}, {and} \bibinfo{person}{Yu Wang}.} \bibinfo{year}{2022}\natexlab{}.
\newblock \showarticletitle{Optimizing Graph-based Approximate Nearest Neighbor Search: Stronger and Smarter}. In \bibinfo{booktitle}{\emph{2022 23rd IEEE International Conference on Mobile Data Management (MDM)}}. IEEE, \bibinfo{pages}{179--184}.
\newblock


\bibitem[\protect\citeauthoryear{Liu, Guo, Liu, Xiao, and Tang}{Liu et~al\mbox{.}}{2023}]%
        {liu2023cmlocate}
\bibfield{author}{\bibinfo{person}{Zhuoqun Liu}, \bibinfo{person}{Fan Guo}, \bibinfo{person}{Heng Liu}, \bibinfo{person}{Xiaoyue Xiao}, {and} \bibinfo{person}{Jin Tang}.} \bibinfo{year}{2023}\natexlab{}.
\newblock \showarticletitle{CMLocate: A cross-modal automatic visual geo-localization framework for a natural environment without GNSS information}.
\newblock \bibinfo{journal}{\emph{IET Image Processing}} (\bibinfo{year}{2023}).
\newblock


\bibitem[\protect\citeauthoryear{Lloyd}{Lloyd}{1982}]%
        {1056489}
\bibfield{author}{\bibinfo{person}{S. Lloyd}.} \bibinfo{year}{1982}\natexlab{}.
\newblock \showarticletitle{Least squares quantization in PCM}.
\newblock \bibinfo{journal}{\emph{IEEE Transactions on Information Theory}} \bibinfo{volume}{28}, \bibinfo{number}{2} (\bibinfo{year}{1982}), \bibinfo{pages}{129--137}.
\newblock
\urldef\tempurl%
\url{https://doi.org/10.1109/TIT.1982.1056489}
\showDOI{\tempurl}


\bibitem[\protect\citeauthoryear{Maddison, Mnih, and Teh}{Maddison et~al\mbox{.}}{2016}]%
        {maddison2016concrete}
\bibfield{author}{\bibinfo{person}{Chris~J Maddison}, \bibinfo{person}{Andriy Mnih}, {and} \bibinfo{person}{Yee~Whye Teh}.} \bibinfo{year}{2016}\natexlab{}.
\newblock \showarticletitle{The concrete distribution: A continuous relaxation of discrete random variables}.
\newblock \bibinfo{journal}{\emph{arXiv preprint arXiv:1611.00712}} (\bibinfo{year}{2016}).
\newblock


\bibitem[\protect\citeauthoryear{Malkov, Ponomarenko, Logvinov, and Krylov}{Malkov et~al\mbox{.}}{2014}]%
        {malkov2014approximate}
\bibfield{author}{\bibinfo{person}{Yury Malkov}, \bibinfo{person}{Alexander Ponomarenko}, \bibinfo{person}{Andrey Logvinov}, {and} \bibinfo{person}{Vladimir Krylov}.} \bibinfo{year}{2014}\natexlab{}.
\newblock \showarticletitle{Approximate nearest neighbor algorithm based on navigable small world graphs}.
\newblock \bibinfo{journal}{\emph{Information Systems}}  \bibinfo{volume}{45} (\bibinfo{year}{2014}), \bibinfo{pages}{61--68}.
\newblock


\bibitem[\protect\citeauthoryear{Malkov and Yashunin}{Malkov and Yashunin}{2018}]%
        {malkov2018efficient}
\bibfield{author}{\bibinfo{person}{Yu~A Malkov} {and} \bibinfo{person}{Dmitry~A Yashunin}.} \bibinfo{year}{2018}\natexlab{}.
\newblock \showarticletitle{Efficient and robust approximate nearest neighbor search using hierarchical navigable small world graphs}.
\newblock \bibinfo{journal}{\emph{IEEE transactions on pattern analysis and machine intelligence}} \bibinfo{volume}{42}, \bibinfo{number}{4} (\bibinfo{year}{2018}), \bibinfo{pages}{824--836}.
\newblock


\bibitem[\protect\citeauthoryear{Matsui, Imaizumi, Miyamoto, and Yoshifuji}{Matsui et~al\mbox{.}}{2022}]%
        {matsui2022arm}
\bibfield{author}{\bibinfo{person}{Yusuke Matsui}, \bibinfo{person}{Yoshiki Imaizumi}, \bibinfo{person}{Naoya Miyamoto}, {and} \bibinfo{person}{Naoki Yoshifuji}.} \bibinfo{year}{2022}\natexlab{}.
\newblock \showarticletitle{ARM 4-BIT PQ: SIMD-Based Acceleration for Approximate Nearest Neighbor Search on ARM}. In \bibinfo{booktitle}{\emph{ICASSP 2022-2022 IEEE International Conference on Acoustics, Speech and Signal Processing (ICASSP)}}. IEEE, \bibinfo{pages}{2080--2084}.
\newblock


\bibitem[\protect\citeauthoryear{Meng, Dai, Yan, Cheng, Liu, Guo, Liao, and Chen}{Meng et~al\mbox{.}}{2020}]%
        {meng2020pmd}
\bibfield{author}{\bibinfo{person}{Yitong Meng}, \bibinfo{person}{Xinyan Dai}, \bibinfo{person}{Xiao Yan}, \bibinfo{person}{James Cheng}, \bibinfo{person}{Weiwen Liu}, \bibinfo{person}{Jun Guo}, \bibinfo{person}{Benben Liao}, {and} \bibinfo{person}{Guangyong Chen}.} \bibinfo{year}{2020}\natexlab{}.
\newblock \showarticletitle{Pmd: An optimal transportation-based user distance for recommender systems}. In \bibinfo{booktitle}{\emph{Advances in Information Retrieval: 42nd European Conference on IR Research, ECIR 2020, Lisbon, Portugal, April 14--17, 2020, Proceedings, Part II 42}}. Springer, \bibinfo{pages}{272--280}.
\newblock


\bibitem[\protect\citeauthoryear{Meta}{Meta}{[n.d.]}]%
        {faissweb}
\bibfield{author}{\bibinfo{person}{Meta}.} \bibinfo{year}{[n.d.]}\natexlab{}.
\newblock \bibinfo{title}{A library for efficient similarity search and clustering of dense vectors.}
\newblock \bibinfo{howpublished}{\url{https://github.com/facebookresearch/faiss}}.
\newblock


\bibitem[\protect\citeauthoryear{Norouzi and Fleet}{Norouzi and Fleet}{2013}]%
        {norouzi2013cartesian}
\bibfield{author}{\bibinfo{person}{Mohammad Norouzi} {and} \bibinfo{person}{David~J Fleet}.} \bibinfo{year}{2013}\natexlab{}.
\newblock \showarticletitle{Cartesian k-means}. In \bibinfo{booktitle}{\emph{Proceedings of the IEEE Conference on computer Vision and Pattern Recognition}}. \bibinfo{pages}{3017--3024}.
\newblock


\bibitem[\protect\citeauthoryear{Peng, Choi, Chan, and Xu}{Peng et~al\mbox{.}}{2022}]%
        {peng2022lan}
\bibfield{author}{\bibinfo{person}{Yun Peng}, \bibinfo{person}{Byron Choi}, \bibinfo{person}{Tsz~Nam Chan}, {and} \bibinfo{person}{Jianliang Xu}.} \bibinfo{year}{2022}\natexlab{}.
\newblock \showarticletitle{Lan: Learning-based approximate k-nearest neighbor search in graph databases}. In \bibinfo{booktitle}{\emph{2022 IEEE 38th international conference on data engineering (ICDE)}}. IEEE, \bibinfo{pages}{2508--2521}.
\newblock


\bibitem[\protect\citeauthoryear{Prokhorenkova and Shekhovtsov}{Prokhorenkova and Shekhovtsov}{2020}]%
        {prokhorenkova2020graph}
\bibfield{author}{\bibinfo{person}{Liudmila Prokhorenkova} {and} \bibinfo{person}{Aleksandr Shekhovtsov}.} \bibinfo{year}{2020}\natexlab{}.
\newblock \showarticletitle{Graph-based nearest neighbor search: From practice to theory}. In \bibinfo{booktitle}{\emph{International Conference on Machine Learning}}. PMLR, \bibinfo{pages}{7803--7813}.
\newblock


\bibitem[\protect\citeauthoryear{Ren, Zhang, and Li}{Ren et~al\mbox{.}}{2020}]%
        {ren2020hm}
\bibfield{author}{\bibinfo{person}{Jie Ren}, \bibinfo{person}{Minjia Zhang}, {and} \bibinfo{person}{Dong Li}.} \bibinfo{year}{2020}\natexlab{}.
\newblock \showarticletitle{Hm-ann: Efficient billion-point nearest neighbor search on heterogeneous memory}.
\newblock \bibinfo{journal}{\emph{Advances in Neural Information Processing Systems}}  \bibinfo{volume}{33} (\bibinfo{year}{2020}), \bibinfo{pages}{10672--10684}.
\newblock


\bibitem[\protect\citeauthoryear{Research}{Research}{2023}]%
        {YandexBenchmark}
\bibfield{author}{\bibinfo{person}{Yandex Research}.} \bibinfo{year}{2023}\natexlab{}.
\newblock \bibinfo{title}{Benchmarks for Billion-Scale Similarity Search}.
\newblock
\newblock
\urldef\tempurl%
\url{https://research.yandex.com/blog/benchmarks-for-billion-scale-similarity-search}
\showURL{%
\tempurl}
\newblock
\shownote{Retrieved June 1, 2023.}


\bibitem[\protect\citeauthoryear{Sablayrolles, Douze, Schmid, and J{\'e}gou}{Sablayrolles et~al\mbox{.}}{2018}]%
        {sablayrolles2018spreading}
\bibfield{author}{\bibinfo{person}{Alexandre Sablayrolles}, \bibinfo{person}{Matthijs Douze}, \bibinfo{person}{Cordelia Schmid}, {and} \bibinfo{person}{Herv{\'e} J{\'e}gou}.} \bibinfo{year}{2018}\natexlab{}.
\newblock \showarticletitle{Spreading vectors for similarity search}.
\newblock \bibinfo{journal}{\emph{arXiv preprint arXiv:1806.03198}} (\bibinfo{year}{2018}).
\newblock


\bibitem[\protect\citeauthoryear{Sarwar, Karypis, Konstan, and Riedl}{Sarwar et~al\mbox{.}}{2001}]%
        {sarwar2001item}
\bibfield{author}{\bibinfo{person}{Badrul Sarwar}, \bibinfo{person}{George Karypis}, \bibinfo{person}{Joseph Konstan}, {and} \bibinfo{person}{John Riedl}.} \bibinfo{year}{2001}\natexlab{}.
\newblock \showarticletitle{Item-based collaborative filtering recommendation algorithms}. In \bibinfo{booktitle}{\emph{Proceedings of the 10th international conference on World Wide Web}}. \bibinfo{pages}{285--295}.
\newblock


\bibitem[\protect\citeauthoryear{Shimomura, Oyamada, Vieira, and Kaster}{Shimomura et~al\mbox{.}}{2021}]%
        {shimomura2021survey}
\bibfield{author}{\bibinfo{person}{Larissa~C Shimomura}, \bibinfo{person}{Rafael~Seidi Oyamada}, \bibinfo{person}{Marcos~R Vieira}, {and} \bibinfo{person}{Daniel~S Kaster}.} \bibinfo{year}{2021}\natexlab{}.
\newblock \showarticletitle{A survey on graph-based methods for similarity searches in metric spaces}.
\newblock \bibinfo{journal}{\emph{Information Systems}}  \bibinfo{volume}{95} (\bibinfo{year}{2021}), \bibinfo{pages}{101507}.
\newblock


\bibitem[\protect\citeauthoryear{Simhadri, Williams, Aum{\"u}ller, Douze, Babenko, Baranchuk, Chen, Hosseini, Krishnaswamny, Srinivasa, et~al\mbox{.}}{Simhadri et~al\mbox{.}}{2022}]%
        {simhadri2022results}
\bibfield{author}{\bibinfo{person}{Harsha~Vardhan Simhadri}, \bibinfo{person}{George Williams}, \bibinfo{person}{Martin Aum{\"u}ller}, \bibinfo{person}{Matthijs Douze}, \bibinfo{person}{Artem Babenko}, \bibinfo{person}{Dmitry Baranchuk}, \bibinfo{person}{Qi Chen}, \bibinfo{person}{Lucas Hosseini}, \bibinfo{person}{Ravishankar Krishnaswamny}, \bibinfo{person}{Gopal Srinivasa}, {et~al\mbox{.}}} \bibinfo{year}{2022}\natexlab{}.
\newblock \showarticletitle{Results of the NeurIPS’21 Challenge on Billion-Scale Approximate Nearest Neighbor Search}. In \bibinfo{booktitle}{\emph{NeurIPS 2021 Competitions and Demonstrations Track}}. PMLR, \bibinfo{pages}{177--189}.
\newblock


\bibitem[\protect\citeauthoryear{Singh, Subramanya, Krishnaswamy, and Simhadri}{Singh et~al\mbox{.}}{2021}]%
        {singh2021freshdiskann}
\bibfield{author}{\bibinfo{person}{Aditi Singh}, \bibinfo{person}{Suhas~Jayaram Subramanya}, \bibinfo{person}{Ravishankar Krishnaswamy}, {and} \bibinfo{person}{Harsha~Vardhan Simhadri}.} \bibinfo{year}{2021}\natexlab{}.
\newblock \showarticletitle{FreshDiskANN: A Fast and Accurate Graph-Based ANN Index for Streaming Similarity Search}.
\newblock \bibinfo{journal}{\emph{arXiv preprint arXiv:2105.09613}} (\bibinfo{year}{2021}).
\newblock


\bibitem[\protect\citeauthoryear{Torralba, Fergus, and Weiss}{Torralba et~al\mbox{.}}{2008}]%
        {torralba2008small}
\bibfield{author}{\bibinfo{person}{Antonio Torralba}, \bibinfo{person}{Rob Fergus}, {and} \bibinfo{person}{Yair Weiss}.} \bibinfo{year}{2008}\natexlab{}.
\newblock \showarticletitle{Small codes and large image databases for recognition}. In \bibinfo{booktitle}{\emph{2008 IEEE Conference on Computer Vision and Pattern Recognition}}. IEEE, \bibinfo{pages}{1--8}.
\newblock


\bibitem[\protect\citeauthoryear{Wang, Liu, Kumar, and Chang}{Wang et~al\mbox{.}}{2015}]%
        {wang2015learning}
\bibfield{author}{\bibinfo{person}{Jun Wang}, \bibinfo{person}{Wei Liu}, \bibinfo{person}{Sanjiv Kumar}, {and} \bibinfo{person}{Shih-Fu Chang}.} \bibinfo{year}{2015}\natexlab{}.
\newblock \showarticletitle{Learning to hash for indexing big data—A survey}.
\newblock \bibinfo{journal}{\emph{Proc. IEEE}} \bibinfo{volume}{104}, \bibinfo{number}{1} (\bibinfo{year}{2015}), \bibinfo{pages}{34--57}.
\newblock


\bibitem[\protect\citeauthoryear{Wang, Lv, Xu, Wang, Yue, and Ni}{Wang et~al\mbox{.}}{2022a}]%
        {Wang2022NHQ}
\bibfield{author}{\bibinfo{person}{Mengzhao Wang}, \bibinfo{person}{Lingwei Lv}, \bibinfo{person}{Xiaoliang Xu}, \bibinfo{person}{Yuxiang Wang}, \bibinfo{person}{Qiang Yue}, {and} \bibinfo{person}{Jiongkang Ni}.} \bibinfo{year}{2022}\natexlab{a}.
\newblock \showarticletitle{Navigable Proximity Graph-Driven Native Hybrid Queries with Structured and Unstructured Constraints}.
\newblock \bibinfo{journal}{\emph{arXiv preprint arXiv:1412.6980}} (\bibinfo{year}{2022}).
\newblock


\bibitem[\protect\citeauthoryear{Wang, Lv, Xu, Wang, Yue, and Ni}{Wang et~al\mbox{.}}{2024}]%
        {Wang2024}
\bibfield{author}{\bibinfo{person}{Mengzhao Wang}, \bibinfo{person}{Lingwei Lv}, \bibinfo{person}{Xiaoliang Xu}, \bibinfo{person}{Yuxiang Wang}, \bibinfo{person}{Qiang Yue}, {and} \bibinfo{person}{Jiongkang Ni}.} \bibinfo{year}{2024}\natexlab{}.
\newblock \showarticletitle{An Efficient and Robust Framework for Approximate Nearest Neighbor Search with Attribute Constraint}. In \bibinfo{booktitle}{\emph{NeurIPS, Accpeted}}.
\newblock


\bibitem[\protect\citeauthoryear{Wang, Xu, Yue, and Wang}{Wang et~al\mbox{.}}{2021}]%
        {Wang2021}
\bibfield{author}{\bibinfo{person}{Mengzhao Wang}, \bibinfo{person}{Xiaoliang Xu}, \bibinfo{person}{Qiang Yue}, {and} \bibinfo{person}{Yuxiang Wang}.} \bibinfo{year}{2021}\natexlab{}.
\newblock \showarticletitle{A Comprehensive Survey and Experimental Comparison of Graph-Based Approximate Nearest Neighbor Search}.
\newblock \bibinfo{journal}{\emph{Proc. {VLDB} Endow.}} \bibinfo{volume}{14}, \bibinfo{number}{11} (\bibinfo{year}{2021}), \bibinfo{pages}{1964--1978}.
\newblock


\bibitem[\protect\citeauthoryear{Wang, Yin, Chen, Yu, Zhou, and Zhang}{Wang et~al\mbox{.}}{2022b}]%
        {wang2022fast}
\bibfield{author}{\bibinfo{person}{Qinyong Wang}, \bibinfo{person}{Hongzhi Yin}, \bibinfo{person}{Tong Chen}, \bibinfo{person}{Junliang Yu}, \bibinfo{person}{Alexander Zhou}, {and} \bibinfo{person}{Xiangliang Zhang}.} \bibinfo{year}{2022}\natexlab{b}.
\newblock \showarticletitle{Fast-adapting and privacy-preserving federated recommender system}.
\newblock \bibinfo{journal}{\emph{The VLDB Journal}} \bibinfo{volume}{31}, \bibinfo{number}{5} (\bibinfo{year}{2022}), \bibinfo{pages}{877--896}.
\newblock


\bibitem[\protect\citeauthoryear{Wang, Liu, Xu, Ke, Wu, and Gou}{Wang et~al\mbox{.}}{2023}]%
        {Wang2023}
\bibfield{author}{\bibinfo{person}{Yuxiang Wang}, \bibinfo{person}{Jun Liu}, \bibinfo{person}{Xiaoliang Xu}, \bibinfo{person}{Xiangyu Ke}, \bibinfo{person}{Tianxing Wu}, {and} \bibinfo{person}{Xiaoxuan Gou}.} \bibinfo{year}{2023}\natexlab{}.
\newblock \showarticletitle{Efficient and Effective Academic Expert Finding on Heterogeneous Graphs through (k,P)-Core based Embedding}.
\newblock \bibinfo{journal}{\emph{{ACM} Trans. Knowl. Discov. Data}} \bibinfo{volume}{17}, \bibinfo{number}{6} (\bibinfo{year}{2023}), \bibinfo{pages}{85:1--85:35}.
\newblock


\bibitem[\protect\citeauthoryear{Xu, Liu, Wang, and Ke}{Xu et~al\mbox{.}}{2022}]%
        {xu2022academic}
\bibfield{author}{\bibinfo{person}{Xiaoliang Xu}, \bibinfo{person}{Jun Liu}, \bibinfo{person}{Yuxiang Wang}, {and} \bibinfo{person}{Xiangyu Ke}.} \bibinfo{year}{2022}\natexlab{}.
\newblock \showarticletitle{Academic Expert Finding via $(k,P) $-Core based Embedding over Heterogeneous Graphs}. In \bibinfo{booktitle}{\emph{2022 IEEE 38th International Conference on Data Engineering (ICDE)}}. IEEE, \bibinfo{pages}{338--351}.
\newblock


\bibitem[\protect\citeauthoryear{Yang, Wang, Tan, and Xiao}{Yang et~al\mbox{.}}{2021}]%
        {yang2021hierarchical}
\bibfield{author}{\bibinfo{person}{Kaixiang Yang}, \bibinfo{person}{Hongya Wang}, \bibinfo{person}{Ming Du1 Zhizheng Wang1~Zongyuan Tan}, {and} \bibinfo{person}{Yingyuan Xiao}.} \bibinfo{year}{2021}\natexlab{}.
\newblock \showarticletitle{Hierarchical Link and Code: Efficient Similarity Search for Billion-Scale Image Sets}.
\newblock  (\bibinfo{year}{2021}).
\newblock


\bibitem[\protect\citeauthoryear{yq}{yq}{2023}]%
        {BREWESS}
\bibfield{author}{\bibinfo{person}{yq}.} \bibinfo{year}{2023}\natexlab{}.
\newblock \bibinfo{title}{BREWESS}.
\newblock
\newblock
\urldef\tempurl%
\url{https://github.com/Lsyhprum/BREWESS.git}
\showURL{%
\tempurl}
\newblock
\shownote{Retrieved June 1, 2023.}


\bibitem[\protect\citeauthoryear{Zhan, Mao, Liu, Guo, Zhang, and Ma}{Zhan et~al\mbox{.}}{2021}]%
        {zhan2021optimizing}
\bibfield{author}{\bibinfo{person}{Jingtao Zhan}, \bibinfo{person}{Jiaxin Mao}, \bibinfo{person}{Yiqun Liu}, \bibinfo{person}{Jiafeng Guo}, \bibinfo{person}{Min Zhang}, {and} \bibinfo{person}{Shaoping Ma}.} \bibinfo{year}{2021}\natexlab{}.
\newblock \showarticletitle{Optimizing dense retrieval model training with hard negatives}. In \bibinfo{booktitle}{\emph{Proceedings of the 44th International ACM SIGIR Conference on Research and Development in Information Retrieval}}. \bibinfo{pages}{1503--1512}.
\newblock


\bibitem[\protect\citeauthoryear{Zhang, Tang, Hu, and Wang}{Zhang et~al\mbox{.}}{2022}]%
        {zhang2022connecting}
\bibfield{author}{\bibinfo{person}{Haokui Zhang}, \bibinfo{person}{Buzhou Tang}, \bibinfo{person}{Wenze Hu}, {and} \bibinfo{person}{Xiaoyu Wang}.} \bibinfo{year}{2022}\natexlab{}.
\newblock \showarticletitle{Connecting Compression Spaces with Transformer for Approximate Nearest Neighbor Search}. In \bibinfo{booktitle}{\emph{European Conference on Computer Vision}}. Springer, \bibinfo{pages}{515--530}.
\newblock


\bibitem[\protect\citeauthoryear{Zhang and He}{Zhang and He}{2019}]%
        {zhang2019grip}
\bibfield{author}{\bibinfo{person}{Minjia Zhang} {and} \bibinfo{person}{Yuxiong He}.} \bibinfo{year}{2019}\natexlab{}.
\newblock \showarticletitle{Grip: Multi-store capacity-optimized high-performance nearest neighbor search for vector search engine}. In \bibinfo{booktitle}{\emph{Proceedings of the 28th ACM International Conference on Information and Knowledge Management}}. \bibinfo{pages}{1673--1682}.
\newblock


\bibitem[\protect\citeauthoryear{Zhu, Zhu, Li, Bi, and Song}{Zhu et~al\mbox{.}}{2019}]%
        {zhu2019accelerating}
\bibfield{author}{\bibinfo{person}{Chun~Jiang Zhu}, \bibinfo{person}{Tan Zhu}, \bibinfo{person}{Haining Li}, \bibinfo{person}{Jinbo Bi}, {and} \bibinfo{person}{Minghu Song}.} \bibinfo{year}{2019}\natexlab{}.
\newblock \showarticletitle{Accelerating large-scale molecular similarity search through exploiting high performance computing}. In \bibinfo{booktitle}{\emph{2019 IEEE International Conference on Bioinformatics and Biomedicine (BIBM)}}. IEEE, \bibinfo{pages}{330--333}.
\newblock


\end{thebibliography}

\end{document}